\documentclass[fleqn,usenatbib]{mnras}

\usepackage{newtxtext,newtxmath}

\usepackage{mathtools}
\usepackage[T1]{fontenc}
\usepackage{ae,aecompl}

\usepackage{url}
\usepackage{xspace}
\usepackage{multirow}
\usepackage{adjustbox}
\usepackage{booktabs}

\usepackage{tikz,xcolor}
\definecolor{lime}{HTML}{A6CE39}
\DeclareRobustCommand{\orcidicon}{%
    \begin{tikzpicture}
    \draw[lime, fill=lime] (0,0) 
    circle [radius=0.16] 
    node[white] {{\fontfamily{qag}\selectfont \tiny ID}};
    \draw[white, fill=white] (-0.0625,0.095) 
    circle [radius=0.007];
    \end{tikzpicture}
    \hspace{-2mm}
}
\newcommand{\orcidMY}{\href{https://orcid.org/0000-0001-8697-2385}{\orcidicon}}
\newcommand{\orcidSO}{\href{https://orcid.org/0000-0002-8301-0604}{\orcidicon}}
\newcommand{\orcidAS}{\href{https://orcid.org/0000-0003-3441-9355}{\orcidicon}}
\newcommand{\orcidBK}{\href{https://orcid.org/0000-0001-6695-1157}{\orcidicon}}

\newcommand{\sok}[3]{\color{black}}
\newcommand{\bk}[4]{\color{blue}}

\newcommand{\anm}{AN~UMa\xspace}
\newcommand{\xmmn}{\textit{XMM-Newton}\xspace}
\newcommand{\msun}{${M}_{\odot}$\xspace}

\newcommand{\msunyr}{\msun~yr$^{-1}$\xspace}

\title[Revisiting a Prototype: \anm]{X-ray and Optical Analysis of the Prototype Polar AN UMa}

\author[S. Ok, M. Yardımcı, B. Kalomeni and A. Schwope]
{
    Samet Ok\orcidSO$^{1, 2}$\thanks{E-mail: \href{mailto:samet.ok@gmail.com}{samet.ok@gmail.com}},
       Melis Yardımcı\orcidMY$^{1}$, Belinda Kalomeni\orcidBK$^{1}$ and Axel Schwope\orcidAS$^{2}$   
       \\
   {\normalsize $^{1}$Department of Astronomy and Space Sciences, University of Ege, 35100, {\.I}zmir, Türkiye}  \\
   {\normalsize $^{2}$Leibniz Institute for Astrophysics Potsdam (AIP), An der Sternwarte 	16, 14482 Potsdam, Germany}
   }

\date{Accepted XXX. Received YYY; in original form ZZZ}

\pubyear{2025}

\begin{document}
\label{firstpage}
\pagerange{\pageref{firstpage}--\pageref{lastpage}}
\maketitle

\begin{abstract}
We present a long-term optical and X-ray photometric study of AN UMa, one of the prototypical polar-type cataclysmic variables, tracing more than 34~years of its accretion history. Observations from both ground-based and space-based facilities have been analysed to investigate state transitions within the system.
Throughout this period, significant changes in the light curve have been observed, corresponding to different mass accretion states. From four years of TESS photometry, we derive a revised photometric period that agrees with the spectroscopic period to within 1.2$\sigma$. These optical observations further suggest switching between two accretion poles.
During intervals of high accretion, dips in the X-ray light curve indicate that the primary accretion pole is obscured by an accretion stream elevated above the orbital plane, a feature also evident in the TESS light curves. Additionally, periodogram analysis reveals a periodicity of $\approx$437 days, which may be related to long-term accretion state changes.
Following a 16-year high state, AN UMa entered two short-lived low states, lasting 180 and 123~days, during which it faded to a magnitude of 19.2, as recorded by the ZTF and ATLAS surveys. Using the system’s low-state brightness and the distance provided by Gaia, we estimate that the system may have a white dwarf with an effective temperature of $\approx$15000~K, and a donor of spectral type M4.7.
This work provides a useful reference for future studies of polars with variable accretion geometries and highlights the importance of long-term, multi-wavelength monitoring in the study of magnetic cataclysmic variables.
\end{abstract}



\section{INTRODUCTION}
Cataclysmic variables (CVs) are short-period interacting binary systems consisting of a mass-accreting white dwarf (WD) and a low-mass donor star. As the massive star in the primordial binaries of CVs evolves, dynamically unstable mass transfer may start under certain conditions. This stage of evolution is known as the common envelope (CE) phase, where the less massive star spiral-in towards the envelope of the giant star \citep[see,][]{paczynski76,taam+78,webbink84,zorotovic+11}. At the end of CE phase, a binary system consisting of a WD in a very tight orbit can form. For an extensive discussion of the evolution of CVs, we refer interested readers to the papers \citet{rappaport+83, patterson84, warner95,howell+01,goliasch+15,kalomeni+16}, and references therein.

CVs can be classified based on their observed physical characteristics. Systems containing highly magnetic WDs are generally called magnetic CVs (mCVs). Polars  ($\sim$10 -- 230~MG) and intermediate polars (IPs) ($\sim$1 -- 10~MG) can be classified as mCVs. Magnetic field strength plays a key role in determining the accretion geometry in these short-period mass-transferring interacting binary systems. In polars, for instance, an accretion disc is not formed, and the accreted material is channelled along magnetic field lines, and relatively less magnetic WDs in IPs result in disrupted accretion discs. In the non-mCVs, on the other hand, the WD magnetic field does not govern the accretion pattern and an accretion disc is formed. 

In addition to orbital period variations, polars exhibit irregular brightness changes, known as high and low states, driven by their variable mass transfer rates.
Low states of polars, therefore, are favourable states to study the stars in the binary \citep[see e.g.][]{szkody+81,liebert+82}. Very rapid variations that were observed in \anm suggested to originate from the base of the accretion column, and later this feature was observed among the other polars \citep[see, ][]{imamura+86,middleditch+97}. 

AN~Ursae~Majoris or \anm is the second polar to be discovered \citep{bond+74} after AM~Herculis. \cite{kukarkin+75} reported \anm to be very short period (0$^\mathrm{d}$.15950) eclipsing nova-like binary system. Strong linear and circular polarisation related to the strong magnetic field on the white dwarf was reported by \citep{krzeminski+77}. These authors nicknamed these systems "polars". Polarisation observations of \anm were interpreted in a one-pole accretion scenario \citep[see e.g.][]{szkody+81, liebert+82}. Its vicinity and brightness led to many photometric, spectroscopic, and polarimetric studies during high and low accretion states \citep[see e.g. ][]{szkody+81,szkody+88,ramseyer+93,bonnet-bidaud+96,ramsay+04b,campbell+08, kalomeni12, harrison+15}.


The \textit{V} band brightness of \anm was observed to vary from 16$^\mathrm{m}$.5 to 19$^\mathrm{m}$ at high and low states, respectively \citep{garnavich+88}. The same study shows that the system reached a brightness of 14$^\mathrm{m}$.5, which they stated as a very high state. In both states, the light curve of \anm has an orbital variability amplitude of about 1 magnitude \citep{liebert+82}. \cite{shugarov78} showed with multi-band photometry that the optical light curve of \anm exhibits a single dip or consecutive hump-like structure in its light curve. However, \cite{ramseyer+93}, based on multi-band photometry, reported two dips or two hump-like features separated by approximately 0.5 phase in the light curve. \cite{mumford77} observed a narrower and deeper minimum-like or eclipse-like dip in one of these humps at the high state than it was observed at the low state. 

The orbital period of \anm was determined to be $P_{\rm orb}$~=~ 0.07975282(4) days by \cite{bonnet-bidaud+96}, using HeII (4686\AA) and H$_{\gamma}$ emission lines. This period was compared with the polarimetric period of 0.07975320(3) days reported by \cite{liebert+82}, revealing a slight discrepancy suggestive of potential asynchronism ($P_{\rm orb}\ne P_{\rm spin}$). Nevertheless, \anm is still classified as a synchronous polar.

The mass and temperature of the WD in the system are proposed to be $0.4-0.6$~\msun and 20000~K \citep{bonnet-bidaud+96}. In a recent study \cite{bonnet-bidaud+15} revised the WD mass as 1~\msun from the $\it B$-$\dot {M}$ relation. The existence of variable and strong emission lines is explained by the mass transfer from the low mass donor star ($M_{\rm don}\approx 0.22-0.24$~\msun) to the magnetic WD accretor \citep{krzeminski+77}. Currently, the parameters of the system's components have not been measured with sufficient sensitivity, especially considering the recent Gaia distance, and the proposed values embrace an extensive range. Furthermore, these values deviate somewhat from the average mass value commonly accepted for CVs \citep{pala+17}, and therefore warrant further investigation.

The magnetic field intensity of \anm was inferred from the identification of cyclotron harmonics in optical and infrared spectra. \cite{cropper+89} deduced a field strength of 35.8~MG while  \cite{campbell+08_B} reported a lower magnetic field of 32.1~MG. They explained this difference by a change in the mass accretion rate. \cite{campbell+08_B} reported an inclination angle ($\it {i}$) of 50$^{\circ}$ and 17$^{\circ}$ magnetic colatitude in agreement with the previous studies \citep[i.e.][]{cropper+89,bonnet-bidaud+96}.

Following AM Her, \anm is the second polar detected to produce soft X-rays via Small Astronomy Satellite~3 (\textit{SAS}~3) \citep{hearn+79}. In \textit{EXOSAT} observation, \anm exhibits a 2-hump-like structure in its X-ray light curve\citep{osborne88}. This light curve also displays a deep eclipse-like dip within one of the humps. The dip is likely due to the occultation of the X-ray emitting region of the column by the accretion stream \citep{ramsay+04b}.  It has also been reported that the system has a blackbody component of $kT_{\rm bb}\approx$23~eV from \textit{ROSAT} observations \citep{ramsay+94}.

 
A substantial body of photometric optical and X-ray data on the system has become available in archival sources, particularly since 2000. One observational data set of the system is present in the \xmmn archive. This observation indicates a change in the system’s accretion flow conditions, rendering it suitable for an orbital period-dependent X-ray analysis, presented here for the first time. We revisit previously reported anomalies in the light curves and re-evaluate the system’s accretion state, interpreting features such as changes in accretion geometry and potential asynchronism in light of contemporary data. This study is intended to serve as a reference for AM Her-type and related systems. In Section~\ref{sec:obs}, we describe the ground- and space-based optical and X-ray observations of \anm. Section~\ref{sec:data_analysis} details our analysis and results. Finally, in Section~\ref{sec:conclusions}, we discuss our findings in the context of the existing literature.

\section{OBSERVATIONS} 
\label{sec:obs}
\subsection{Optical Observations}
\label{sec:optical_obs}
Polars are known to be very strongly variable due to orbital variability, projection, and occultation effects and due to long-term accretion rate changes. Thus, regular follow-up observations are useful to separate various effects from each other. Besides periodic orbital variations, polars may show variations on short and long-time scales caused by accretion rate changes and donor star activity. Hence, this study includes follow-up and nightly observations to study any possible short - and long-term variations of \anm. The data from the various observatories and databases used in this study use different time systems (e.g. JD, MJD, BJD, or HJD). To provide integrity, all dates are converted to Barycentric Julian Date (BJD) thanks to using the \textit{astropy}\footnote{\url{https://www.astropy.org/}} package. To compare \anm's past observations to each other, a consistent and precise orbital period is essential.

The most recent ephemeris in the literature was noted in \citet{bonnet-bidaud+96} as:
%
$T_{0}(\rm HJD)=2443190.9921(2)+0.07975282(4) \times \textit{E}.$ 
%
This ephemeris was created by combining two different observational data sets in the related study. The $T_{0}$ obtained from polarimetry corresponds to the linear pulse regarded as the magnetic pole at the limb of the white dwarf and obtained by \cite{liebert+82}, whereas the orbital period has been obtained from spectroscopy by \citet{bonnet-bidaud+96}. The spectroscopic observations used to compute the relevant period date back to 1979. Considering the period error ($\delta$P~$=4\times$10$^{-8}$ day) and the 46 years after, the cumulative period error indicates that a discrepancy of around $\approx$740 seconds may exist today. The difference equates to $\approx$10\% of the orbital period. In order to compare light curves from different times, it is necessary to improve this period. In Sec.~\ref {sec:tess}, we improved this period further with new photometric observations.

We adopt the polarimetric $T_{0}$ due to its physical significance in describing the magnetic accretion geometry and its widespread use in the literature on \anm. Consequently, we applied a time-system transformation to polarimetric $T_{0}$ (HJD~2443190.9921(2)) and converted it into BJD format as BJD~2443190.9926 and employed it in all folded light curves for this study. 

\begin{table}
\centering
\large
\caption{Observation log of \anm. See the text for details.}
\resizebox{\columnwidth}{!}{
\begin{tabular}{lccccc}
\hline
\hline
\textbf{Observatory} & \textbf{Filter} & \textbf{Obs. Starting} & \textbf{Obs. Ending} & \textbf{Duration (d)}  \\
\hline
AAVSO & B  & 19 Jan 2008   & 14 Jun 2014   & 2338.40   \\
AAVSO & V  & 23 Feb 1999   & 24 Apr 2023   & 8531.66    \\
AAVSO & R  & 05 Apr 2005   & 30 Mar 2008   & 1089.95    \\
AAVSO & I  & 08 Apr 2014   & 14 Jun 2014   & 67.52      \\
ASAS-SN & bc-bd & 20 Nov 2013 & 30 Nov 2018   & 1836.04  \\
ATLAS & o  & 08 Dec 2015   & 04 Jul 2023   & 2764.75    \\
ATLAS & c  & 21 Nov 2015   & 21 Jun 2023   & 2799.66    \\
CRTS  & V  & 08 Dec 2005   & 08 Jun 2013   & 2738.76    \\
ZTF   & i  & 06 Apr 2018   & 01 Mar 2022   & 1424.87    \\
ZTF   & g  & 25 Mar 2018   & 28 May 2023   & 1890.02    \\
ZTF   & r  & 06 Apr 2018   & 28 May 2023   & 1878.06    \\
TESS-Sector 21  &  - & 21 Jan 2020   & 18 Feb 2020   & 28 \\
TESS-Sector 48  &  - & 28 Jan 2022   & 26 Feb 2022   & 29  \\
TESS-Sector 75  &  - & 30 Jan 2024   & 26 Feb 2024   & 27 \\
\hline
\textbf{Observatory} & \textbf{Filter} & \textbf{Date}        & \textbf{Instrument}  & \textbf{Exposure (s)} \\
\hline
TUG T60   & V,R   & 2016--2019    & ProLine PL3041 & 60/60    \\
TUG T100  & V,R   & 13 Jun 2016   & SI 1100     & 100/100  \\
TUG T100  & V,R   & 14 Jul 2016   & SI 1100     & 60/50    \\
TUG T100  & V,R   & 16 Jul 2016   & SI 1100     & 60/50    \\
TUG T100  & V,R   & 20 Feb 2017   & SI 1100     & 50/50    \\
TUG T100  & V,R   & 07 Jan 2018   & SI 1100     & 40/40    \\
TUG T100  & R     & 11 Feb 2018   & SI 1100     & 50       \\
TUG T100  & V,R   & 07 Jun 2018   & SI 1100     & 50/50    \\
\hline
Swift    & Photon Counting & 31 May 2007 & XRT     & 4129    \\
Swift    & Photon Counting & 12 Jul 2007 & XRT     & 6934    \\
Swift    & Photon Counting & 28 Feb 2008 & XRT     & 2423    \\
Swift    & Photon Counting & 16 Apr 2008 & XRT     & 2556    \\
Swift    & Photon Counting & 22 Apr 2008 & XRT     & 4117    \\
XMM & THIN1       & 01 May 2002  & EPIC-PN         & 6416    \\
XMM & THIN1       & 01 May 2002  & EPIC-MOS1/MOS2  & 7222    \\
XMM & V,UVW1,UVW2 & 01 May 2002  & OM      & 1301/2500/3000 \\
\hline
\end{tabular}
}
\label{tab:obslog}
\end{table}

\begin{figure*}
    \centering
    \includegraphics[width=0.95\linewidth]{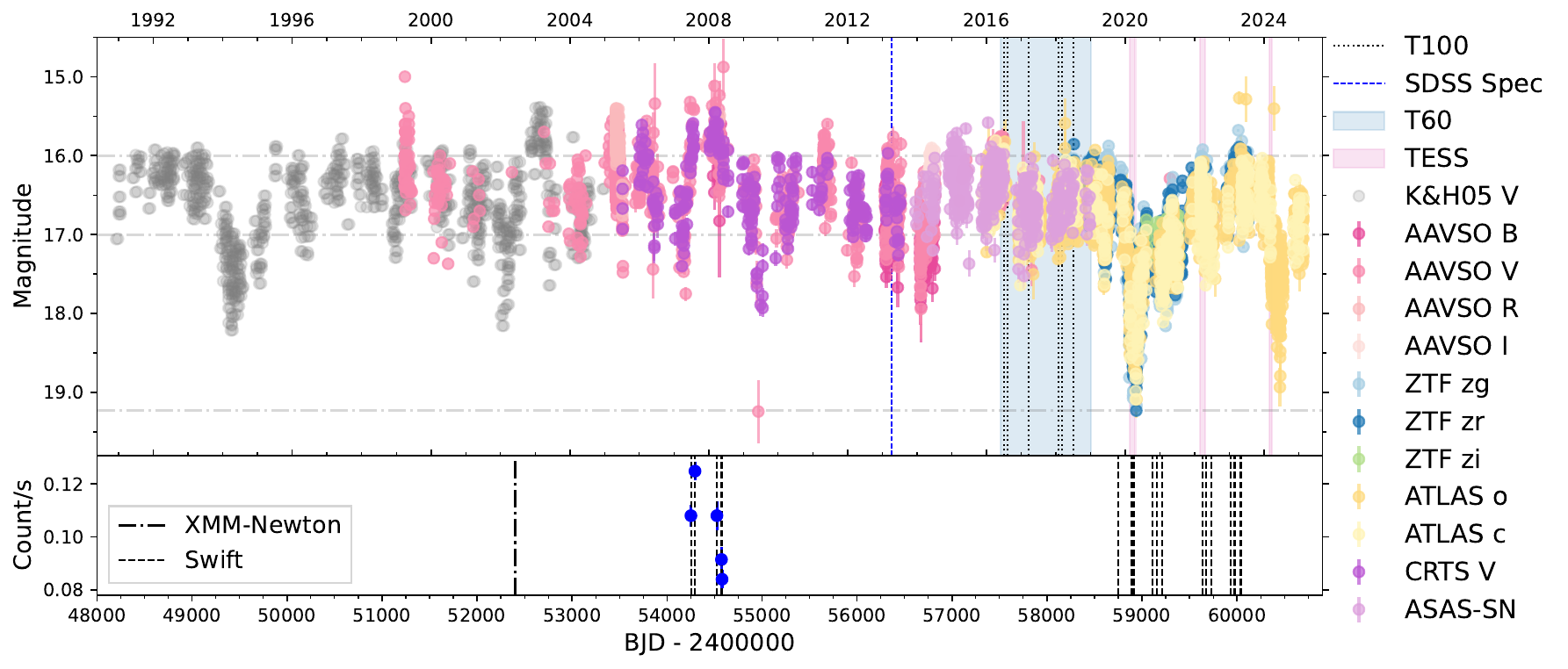}
    \caption{The long-term light curve with the observations of AAVSO, ZTF, ATLAS, CRTS, RoboScope (K\&H05), and ASAS-SN surveys and the information of the X-ray missions. The horizontal grey lines at 16 and 17 magnitudes show the high state where \anm spends the most time, whereas the line at 19.23 magnitude refers to the low state. The dashed lines with mean count values (blue points) show the analysed Swift observations. The remaining vertical dashed lines in the lower panel, spread between 2019 and 2024, indicate the X-ray observations of \anm conducted by the Swift observatory, during which \anm was undetected.}
    \label{fig:lc_ztf+atlas+crts}
\end{figure*}

\subsubsection{AAVSO Data}
The long-term observations of the American Association of Variable Star Observers (AAVSO) extend to the interval of nearly 24 years for the \textit{V}-band (see Tab.~\ref{tab:obslog}). We present AAVSO archival data analyses of \textit{B}, \textit{V}, \textit{R} and \textit{I}-bands from 1999 to 2023 (see Section~\ref{sec:long_term_var}). Although the \textit{V}-band data of AAVSO have the longest duration, the observations do not cover the deep minimum (low state) seen in Fig.~\ref{fig:lc_ztf+atlas+crts}. These and other long-term data are available in the AAVSO Variable Star Index\footnote{\url{https://www.aavso.org/vsx/index.php?view=search.top}} (VSX) catalogue.

\subsubsection{ASAS-SN Observations}

All-Sky Automated Survey for Supernovae \citep[ASAS-SN;][]{shappee+14,kochanek+17} is another survey that had \anm data. \anm has been observed by $bc$ and $bd$ cameras for 5 years at the Hawaii observatory (see Tab.~\ref{tab:obslog}). The light curve displays a flat profile, indicating an absence of significant brightness variations or discernible accretion state transitions.

\subsubsection{ATLAS Observations}
Asteroid Terrestrial-impact Last Alert System \citep[ATLAS;][]{tonry+18} has provided observations of \anm between November 2015 and January 2025. 
Observations were performed by broad passbands of \textit{o} and \textit{c}. We downloaded the data from the interactive 
\textit{ATLAS} database~\footnote{\protect\url{https://fallingstar-data.com}} by using forced photometry (the date of the data request: 26 January 2025). Fig.~\ref{fig:lc_ztf+atlas+crts} shows the long-term brightness behaviour of \anm with other observations. The brightness varies between $15.2 - 19.2$ mag. 
The light variation of ATLAS shows two low states between 2458850 < BJD < 2459030 (lasted 180 days) and between 2460362 < BJD < 2460490. However, the total length of the second low state remains ambiguous due to a gap of around 100~days with no observations covering the end of the minimum. 
The magnitudes reach the lowest value 90~days later in both states. The first deep low state was also obtained from the ZTF survey.

\subsubsection{CRTS Observations}
Catalina Real-Time Transient Survey \citep[CRTS;][]{drake+09} presents 7.5 years of data for \anm. 
Tab.~\ref{tab:obslog} includes observational information of the CRTS \textit{V} band.
During the observation period, the magnitude difference is 2.5 mag. The deepest minimum point of 18 mag is around at BJD~$ \approx 2455005.66 $ (see Section~\ref{sec:long_term_var}). The minimum has also been observed on AAVSO observations that overlapped with CRTS.

\subsubsection{Gaia Observations}
In \textit{Gaia} DR3 \citep{gaia+21} \anm is identified by ID 782275761222286720
 with the sky coordinates of DEC2000=166.10661 deg and RA2000=45.05376 deg. The mean brightness of the object is $16.56 \pm 0.02$, $16.70 \pm 0.06$, and $16.23\pm0.05$ in the \textit{G}, \textit{$G_{\rm BP}$}, and \textit{$G_{\rm RP}$} passbands, respectively. \textit{Gaia} measured the parallax of \anm as $2.986\pm0.056$~mas. We used the geometric distance ($r_{\rm geo}$) to the system of $334^\text{+15}_\text{\,-15}$\,pc, as determined by \cite{bailer-jones+21}. 

\subsubsection{RoboScope Observations}
\anm observed between $1990-2004$ as presented in \citet{kafka+05}. They reported the observations were made with an automated 41~cm telescope in central Indiana known as \textit{RoboScope}. We imported the data from Fig.~8 of the paper using \textit{WebPlotDigitizer} v3.4 (beta)\footnote{\protect{\url{https://web.eecs.utk.edu/~dcostine/personal/PowerDeviceLib/DigiTest/index.html}}} and presented them with grey circles in Fig.~\ref{fig:lc_ztf+atlas+crts} as K\&H05~\textit{V}. 

\subsubsection{TESS Observations}
The Transiting Exoplanet Survey Satellite (TESS) is a space-based telescope \citep{ricker+14a} designed to acquire high-precision time series for individual sources throughout the whole sky.
\anm has observations in three TESS Sectors, 21, 48 and 75, respectively. We downloaded all data from the Mikulski Archive for Space Telescopes database\footnote{\protect\url{https://mast.stsci.edu}}. 
TESS observation intervals are represented in the upper panel of Fig.~\ref{fig:lc_ztf+atlas+crts} as vertical pink bands.
We used pipeline-produced PDCSAP flux in our analyses. The data was obtained with 120-second time bins. The gaps observed across all three sectors correspond to intervals when data acquisition was interrupted due to technical anomalies.

\subsubsection{TUG Observations}

We performed optical photometric observations of \anm with 1~m (T100) and 60~cm (T60) robotic telescopes at T\"{U}B\.{I}TAK National Observatory (TUG) in Türkiye (see Tab.~\ref{tab:obslog} for observation logs). 
Fig.~\ref{fig:lc_ztf+atlas+crts} illustrates the position of TUG observations on the light curve for all missions examined in this work, with the observation period of T60 indicated in a blue band and T100 represented by dotted lines.

\begin{figure}
\centering
   \includegraphics[width=0.85\linewidth]{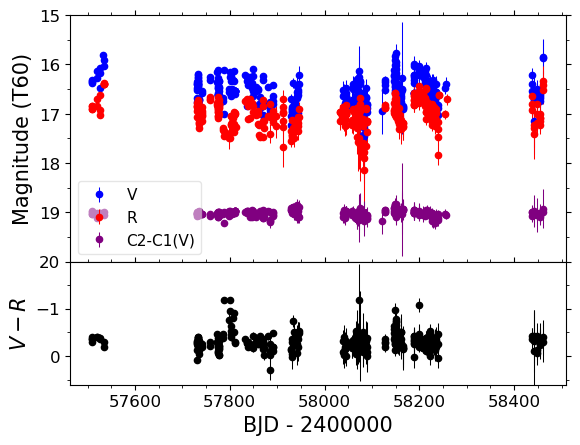}
   \caption{Multiband observations of T60 for \anm. The colours mentioned herein are as follows: \textit{V} (blue), \textit{R} (red), C2-C1 difference of comparison stars in \textit{V}-band (purple) and $V-R$ colour (black). The offset value of C2-C1 is 18.5.}
    \label{fig:t60_lc+color}
\end{figure}

T60 observations cover May 2016 -- December 2018 at Bessel $\it{V}$- and $\it{R}$-bands with integration times of 60~s. 
Differential photometry was performed concerning two nearby stars at 2MASS~J11040840+4502521 and 2MASS~J11040310+4500205 with \text{$\it{V}$-band} magnitudes of $15.608\pm0.075$ and $16.209 \pm0.086$~mag, respectively. The light curves and colour diagram of T60 are seen in Fig.~\ref{fig:t60_lc+color}. 

T100 observations were performed with Bessel $\it{V}$- and $\it{R}$-filters with 2Kx2K SI 1100 CCD from June 2016 -- June 2018 for 6 and 7~nights for $\it{V}$-and $\it{R}$-bands, respectively (see in Fig.~\ref{fig:lc_ztf+atlas+crts}). Exposure time was selected between 40~s and 100~s according to the atmospheric conditions. Data reduced with image reduction program (IRAF) \citep{tody93} and AstroImageJ \citep{collins+17}.

\subsubsection{ZTF Observations}

ZTF \citep[The Zwicky Transient Facility;][]{masci+19} is a time domain survey with an extremely wide field camera using the 1.2~m telescope in Palomar Observatory. \anm has been observed by three filters, \text{ZTF-\textit{i}}, \text{ZTF-\textit{g}}, and \text{ZTF-\textit{r}} (see Tab.~\ref{tab:obslog}). 
The system magnitudes change between 15.7 and 19.2~mag during 2458203~<~BJD~<~2460093 (see Fig. \ref{fig:lc_ztf+atlas+crts}). This low state lasts 180 days between 2458850~<~BJD~<~2459030, with a pronounced period of decrease and increase.

\subsection{X-ray Observations}

\subsubsection{XMM-Newton Observations}

\begin{figure}
	\centering
	\includegraphics[width=0.98\hsize]{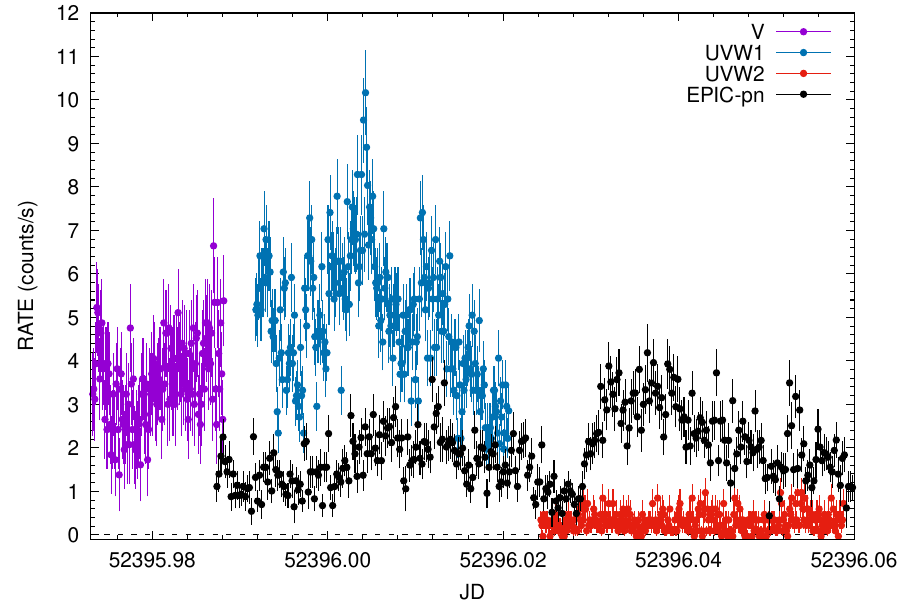}
	\caption{EPIC-pn and OM light curves of \anm.} 
	\label{fig:anuma_om_xray}
\end{figure}

\begin{figure}
\centering
\includegraphics[width=0.98\hsize]{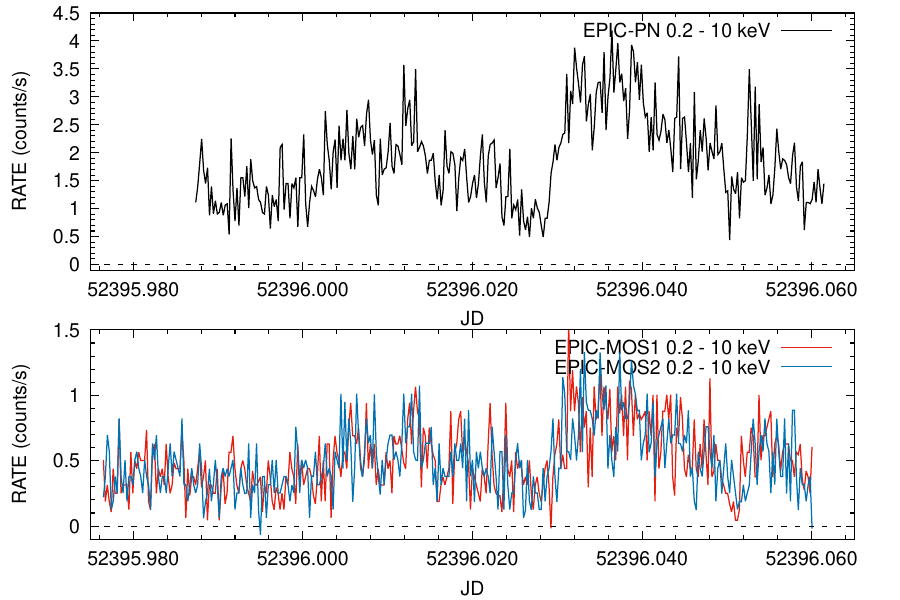}
\caption{\xmmn light curve of \anm in original time series.} 
\label{f:anuma_fold_xray}
\end{figure}

The \xmmn satellite made one observation (ID~0109461701) of \anm on 
5 May 2002 and lasted 7.2~ks\footnote{\url{https://nxsa.esac.esa.int/nxsa-web/}}. The observation covered one orbital cycle ($\sim$1.04). EPIC-pn and MOS instruments were operated in small window mode with a thin filter.
In addition to EPIC-pn and EPIC-MOS cameras, \xmmn observed \anm simultaneously with Optical Monitor (OM) in fast mode with V, UVW1, and UVW2 filters (Fig.~\ref{fig:anuma_om_xray}). The original time series of light curves obtained from EPIC-MOS and EPIC-PN instruments are shown in Fig.~\ref{f:anuma_fold_xray}.

\subsubsection{Swift/XRT Observations}

\anm was observed between 2007 and 2022 by the Swift \citep{gehrels+04} satellite\footnote{\url{https://heasarc.gsfc.nasa.gov/cgi-bin/W3Browse/w3browse.pl}}. Among these observations, only five were suitable for spectral analysis due to the high S/N ratio. We used Obs ID: 00036996001, 00036996003, 00036996005, 90035001, and 90035003 X-ray Telescope (XRT) instrument observations. We used the data from the photon counting ($pc$) mode. Tab.~\ref{tab:obslog} gives detailed information about these observations.

The observations were reduced with the \textit{xrtpipeline} task in \textit{HEASoft} \citep{blackburn+95}. The spectra were obtained from the difference of the source and background apertures with the {\tt XSELECT} command in \textit{HEASoft}.

%
\section{DATA ANALYSIS AND RESULTS}
\label{sec:data_analysis}
\subsection{Optical Photometry}
\label{sec:optical_photometry}
This section presents the analysis and findings of the long-term observations of sky surveys, AAVSO, ASAS-SN, ATLAS, CRTS, T60, and ZTF observatories. Barycentric Julian Date (BJD) was adopted during our analyses, and time corrections were made accordingly.

\subsubsection{TESS Observations}
\label{sec:tess}

\begin{figure*}
    \centering
    \includegraphics[width=1.0\linewidth]{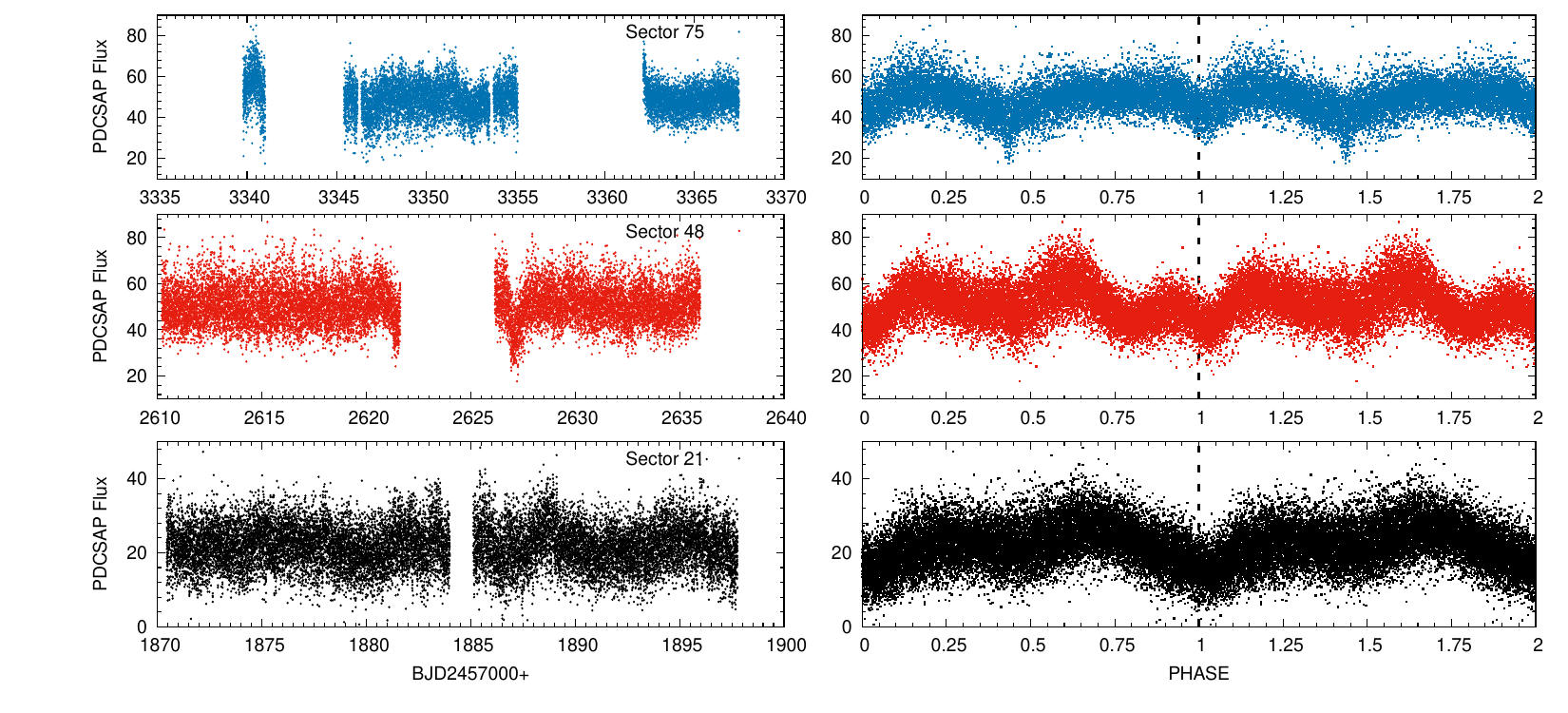}
    \caption{\textit{Left panel}: TESS light curves of \anm in original time series. \textit{Right panel}: Folded light curve of \anm according to ephemeris given in Eq.~\ref{e:eph}.}
    \label{f:tess}
\end{figure*}

TESS has yielded a substantial dataset covering successive orbital cycles on \anm during various mass accretion states. These observations provide a data collection that illustrates various light change behaviours identified in \anm over the years. Each sector reveals a behaviour of a distinct characteristic process, and all are morphologically dissimilar from one another. In analysing this data, we want to understand the circumstances revealed by the X-ray data and explain the origins of the previously indicated changes.

Sector 21 corresponds to the first dataset collected by the TESS (see Fig.~\ref{f:tess}). Sector 21 was obtained beginning the first deep low state detected by the ZTF and ATLAS surveys, and 50\% of this sector was obtained while the system was in a deep low state. The light curve displays a shallow minimum positioned (between phases 0.20 - 0.70) in the centre (0.45 phase) of a prominent hump-like structure, together with the minima between these humps which imply two minimum structures. 

Sector 75 was obtained at the beginning of the second low state, observed by ATLAS, and has morphological resemblances to Sector 21. Half of this observation was obtained inside the second low state, while the other half includes the transition to it. Sector 75 has a sole distinction that is a shallow minimum in the primary hump-like structure appears as a narrow, deep, eclipse-like formation (at phase 0.43) in this sector (see Fig.~\ref{f:tess}). A structure mimicking an eclipse has been first noted by \citet{mumford77}. The eclipse may be associated with this structure. 

Sector 48 exhibits a big difference characterised by three minima that display distinctly different behaviours compared to the others. This sector was obtained in a high accretion state. The shallow minimum observed in the other light curves is also evident here. The deepest point of the minimum occurs within the 0.45 - 0.5 phase interval. The previously referenced mini hump is situated within the phase range of 0.8 to 1.05, with its centre at 0.93 phase. (see Fig.~\ref{f:tess}).

The beginning of the main hump or first dip is evident throughout all three sectors. We used this dip structure to improve the period and test the possible asynchronism previously proposed for the object. We applied a Gaussian fit to this part of the light curve for all sectors. For these fits, the data set was restricted only the data points included around the centre of the dip for $\pm$ 0.125 phase units according to the orbital period for Sector~21, $\pm$ 0.1 phase units for Sector~48, and $\pm$ 0.1 for Sector~75, respectively. We performed our fits with the \textit{lmfit} package in \textit{Python}.

\begin{figure}
    \centering
\includegraphics[width=1.00\linewidth]{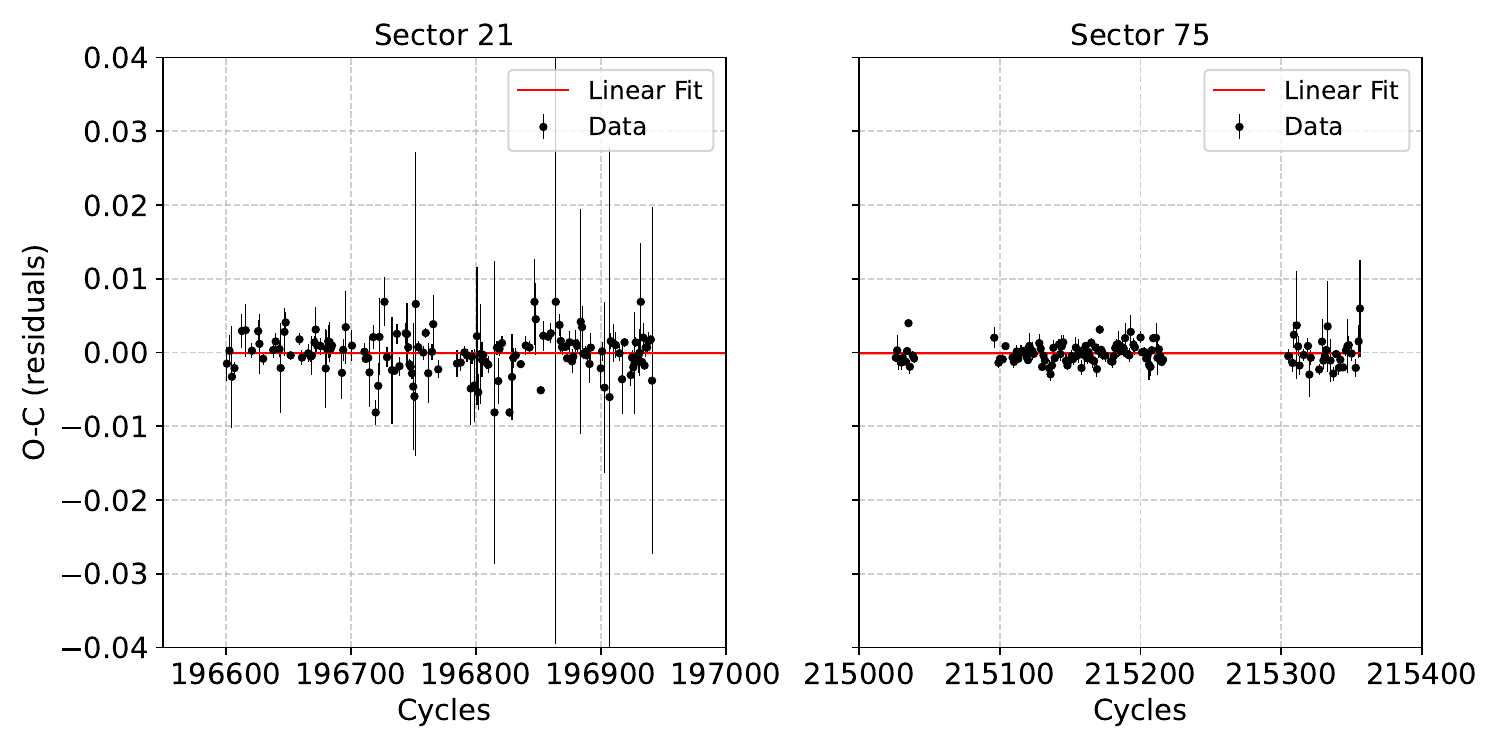}
    \caption{Residuals of the linear regression. The regressions for Sector 21 and Sector 75 are plotted separately for clarity.}
    \label{f:oc}
\end{figure}

We experienced the challenge of modelling the phase interval in Sector~48. Especially in this part, the dip shape was formed by only 4 or 5 data points. In the fit experiments, we observed that the data points were fitted with a certain orientation depending on the phase interval selection. Therefore, we excluded Sector 48 from this calculation and applied the fit only to data points of Sector~21 and Sector~75. We used \citet{liebert+82}'s BJD-converted polarimetric $T_{0}$ as the starting point for the linear regression.
The period we calculated as a result of the weighted linear regression and we applied to the minimum times ($T_{\rm 0, dip}$) obtained from the beginning of the main humps derived as
\begin{equation}
T_{\rm 0}(\text{BJD}) = 2443190.9926(2) + 0.079752867(12) \times E
\label{e:eph}
,\end{equation}
where the numbers in parentheses give the uncertainties in the last digits. The residuals obtained from linear regression are shown in Fig.~\ref{f:oc}. We obtained the photometric period in the order of 10$^{-9}$ day and more precise than the study of \cite{bonnet-bidaud+96}. The difference between this period and the period from \cite{bonnet-bidaud+96} is $\Delta P\approx4.7\times$10$^{-8}$ days. The spectroscopic and photometric periods only differ by 1.2 sigma.  With our TESS data, we do not have the sensitivity to directly demonstrate asynchronism that may be in the system, even if it is to a much smaller degree than suggested by this result. In this respect, we can only say that the system has had a relatively stable orbital period since 1996. In other asynchronous polars with a similar orbital period in CD~Ind \citep{littlefield+19} or the near-synchronous system V1432~Aql \citep{staubert+03} the order of asynchronism is much higher than this difference and comparable with the orbital period (\text{>0.01}$\times~P_{\rm orb}$).

\subsubsection{TUG T100 Observations}
\label{sec:t100}

\begin{figure}
\centering
    \includegraphics[width=0.95\linewidth]{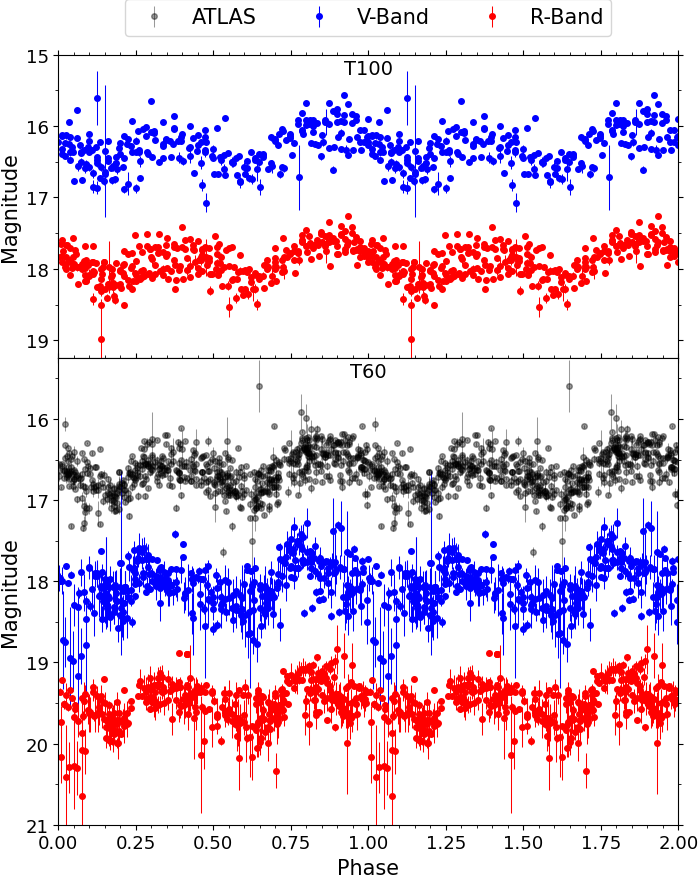}
    \caption{Folded light curves for the TUG T100 (upper panel) and T60 (lower panel) telescopes are shown for the $\it{V}$- and $\it{R}$-bands. The T60 data are presented alongside ATLAS observations for the time interval $2757600~\leq$BJD$\leq~2758400$, corresponding to the same observational epoch. To facilitate comparison between data sets, constant offsets have been applied to some magnitudes: +1.0 for the R-band of T100, +1.5 for the V-band of T60, and +2.5 for the R-band of T60.} 
\label{fig:t100+t60_phase}
\end{figure}

\anm was observed nightly with the TUG T100 telescope, in addition to the archival observations presented in this study. These observations were conducted over six and seven nights in the $\it{V}$ and $\it{R}$ bands, respectively. Double-hump structures are clearly visible in the folded light curves, exhibiting reduced scatter. A comparison between the T60 and T100 observations is possible due to their overlapping time intervals, during which the system remained in the same mass accretion state. Consequently, the folded light curves from both telescopes display similar characteristics, with minima occurring at phases 0.25 and 0.75, based on the photometric period presented in Eq.~\ref{e:eph} (see Fig.~\ref{fig:t100+t60_phase}). The T100 data reveal a pronounced double-hump structure, particularly in the $\it{R}$ band, where the higher signal-to-noise ratio enhances the sharpness of the features.

\subsubsection{ATLAS Observations}

\begin{figure}
    \centering    \includegraphics[width=0.98\linewidth]{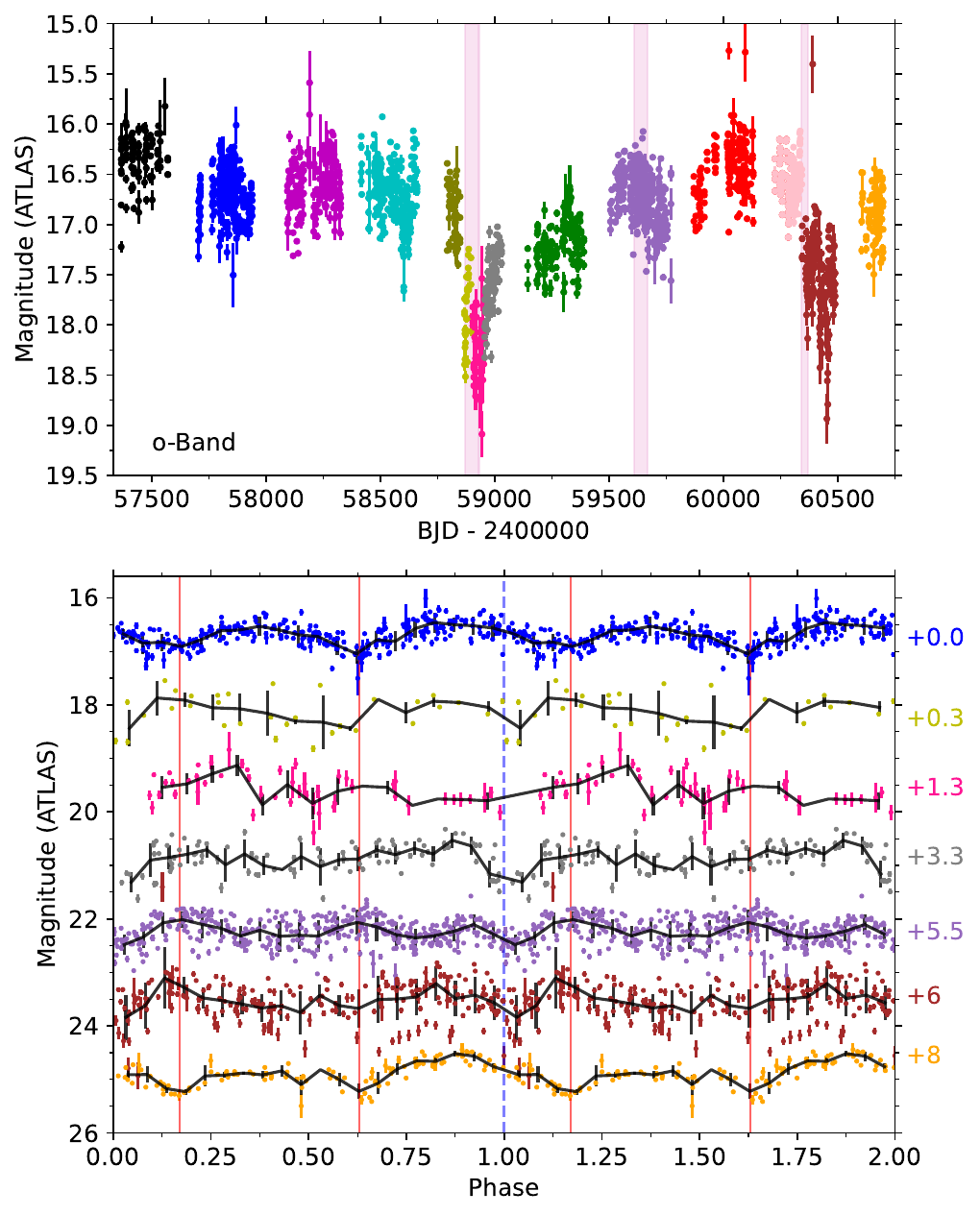}
    \caption{\textit{Upper panel}: ATLAS-\textit{o} light variation of \anm. The pink bands display the TESS observations for Sector~21, 48, and 75. \textit{Lower panel}: 7 folded light curves of \anm. The red solid lines in the lower panel focus on the deviations in the minima or hump-like structures. The colour code is the same for both panels, and the offset values are on the outside of the graph. From the consecutive colour observations displaying the same morphological characteristics, only a single representative folded light curve has been included.
    }
    \label{fig:atlas_8part}
\end{figure}

ATLAS contains ten observational data sets in its archive; however, we divided the light curve into 14 parts to study the light change behaviour according to the system's accretion state. The top panel of Fig.~\ref{fig:atlas_8part} presents the long-term variability of the \text{ATLAS-\textit{o}} data, with each segment coded by a distinct colour-black, blue, magenta, cyan, olive, yellow, deep pink, grey, green, purple, red, pink, brown, and orange-from left to right.

The behaviour of the light variations can be summarised in four stages: (1) a steady phase preceding the low state (black, blue, magenta, cyan, and olive), (2) a deep minimum phase (yellow, deep pink, and grey), (3) a recovery phase (green and purple), and (4) a return to stage (1) (red and pink), with further repetitions (brown and orange).

Every coloured data set is folded according to Eq.~\ref{e:eph}, displaying the transition of the minima into hump-like formations. The lower panel of Fig.~\ref{fig:atlas_8part} displays 7 folded light curves chosen from 14 segments based on their quality and visible distinct features. The data were binned using a phase interval of 0.05; however, for data sets with insufficient data points (olive, yellow, and deep pink), binning was performed using a 0.07 phase interval. The error bars of the binned curves were determined from the standard deviation of the magnitudes within each bin.

The four stages mentioned above are also consistent in the folded light curves. Accordingly, we present only a representative sample for each group of similar morphology. During Stage (1), minima are observed within the phase intervals $0.14$-$0.21$ and $0.58$-$0.65$, as derived from the original, unbinned light curves (see blue light curve in lower panel at Fig.~\ref{fig:atlas_8part}).
However, in Stage (2), this interval was further subdivided into three parts---yellow, deep pink, and grey---allowing the changes during the abrupt decline to be observed more distinctly. 

During Stage (3), the green and purple folded light curves exhibit consistent behaviour, showing two apparent hump-like structures where the minima disappear. Here, the purple one is shown as an example. Additionally, the red, pink, and brown light curves show similar behaviour, notably the absence of a second hump; the brown one was selected as a sample for this case. Finally, the orange folded data display two minima, similar to those seen in the blue data set.

In the initial five observations, the light curves display a dip around phase 0.2. A hump structure, thought to be associated with mass accretion onto the secondary accretion column, occasionally appears in this region. Notably, across these five segments, a hump structure is evident between phases 0.65 and 1.20, and the light curve exhibits a double-hump morphology, consistent with TUG observations. Following the initial transition to a low state, the dip at phase 0.2 is replaced by a hump structure, which persists until the subsequent low state. During this period, the original dip shifts towards phase 0.0. We highlight that, after the second short-lived low state, the final observational segment once again shows the emergence of a dip structure near phase 0.2. 

Although each segment spans several months and consists of sparse data points, potentially masking short-term accretion variability, the overall mass accretion dynamics appear to change systematically following each low-state transition during the ATLAS survey.

\subsubsection{Accretion States of \anm}
The behaviour of the accretion states is the most prominent feature in the long-term light variation of \anm. These states are indicated by three grey horizontal dash-dotted lines in Fig.~\ref{fig:lc_ztf+atlas+crts}. The figure shows that the brightness of \anm predominantly remains between 16 and 17 magnitudes, referred to as the high state. The variation amplitude ($\approx$ 1 mag) is attributed to orbital modulation, as also observed in nightly photometric light curves. Transitions to a very high state, where the brightness exceeds 16 mag, are rare. The low state, in contrast, manifests at brightness levels 1-2 magnitudes fainter than 17 mag. Variability within the low state may indicate the presence of an intermediate state between 17 and 18 mag. A transition to a 19.2 magnitude initially reported by \citet{garnavich+88} as a low state was detected twice by the ATLAS survey. Between 1990 and 2025, only seven low states exceeding 18 magnitudes were recorded for \anm. These times are listed in the spectral energy distribution section (see Sec.~\ref{s:sed}).

\begin{figure}
\centering
\includegraphics[width=0.99\linewidth]{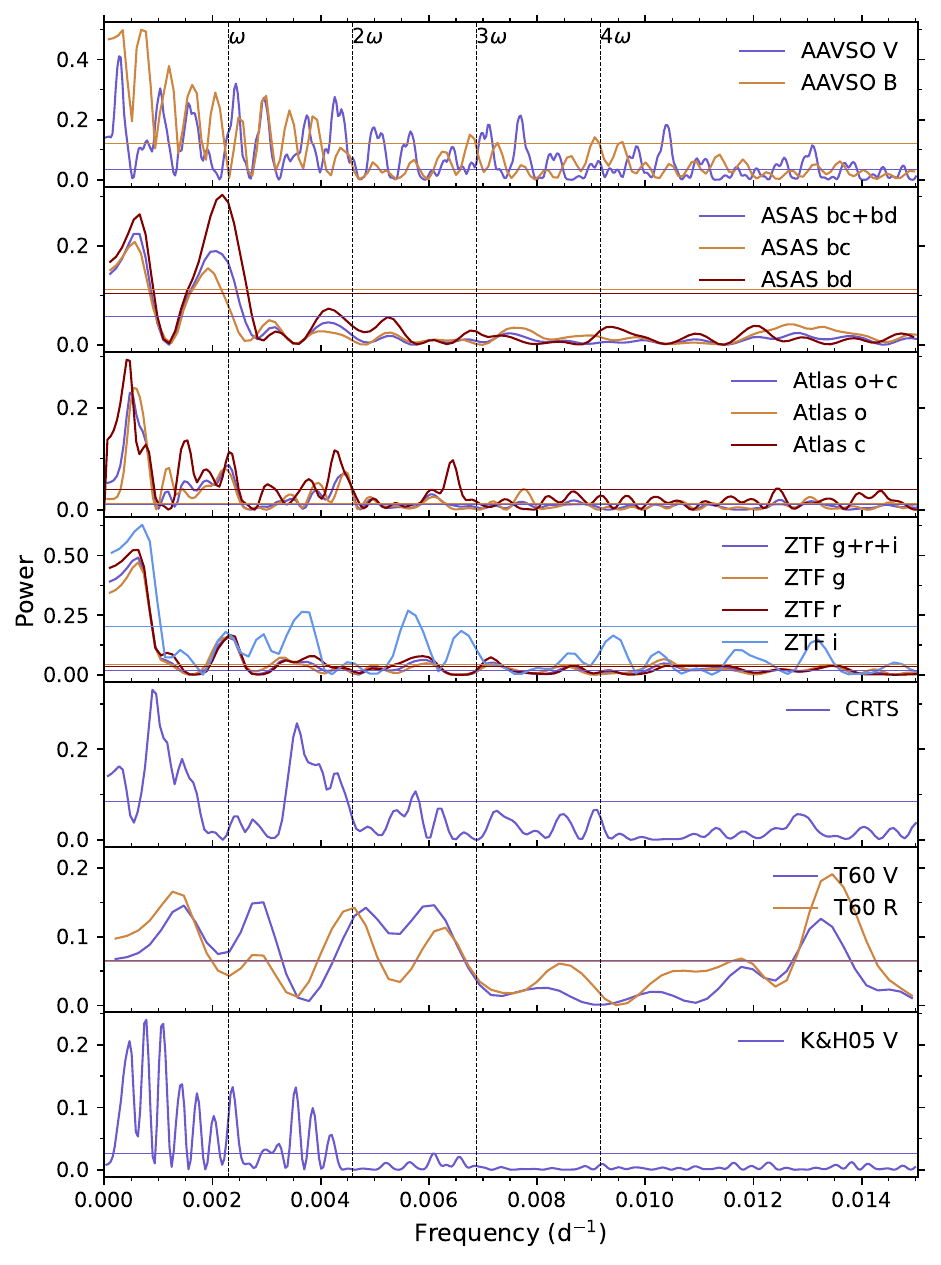}
\caption{The power spectra of \anm. Dotted vertical lines show the locations of $\omega=0.00229$~d$^{-1}$ ($\approx$437~days) and its harmonics. Horizontal lines indicate 99\% confidence levels.}
    \label{fig:freq_analysis}
\end{figure}

\subsubsection{Long-term Variability}
\label{sec:long_term_var}

This section evaluates potential long-term variability in \anm using 34~years of observational data, guided by existing literature recommendations. Specifically, we investigate whether changes occurring over several hundred days may be associated with transitions between accretion states.

We conducted frequency analyses on individual filter data from each observatory, as well as on combined data sets. The photometric frequency analysis was performed using the Lomb–Scargle periodogram \citep{lomb76,scargle82,vanderplas18}, as implemented in the \texttt{astropy} package in \texttt{Python}. Identifying the peak signal is a crucial step in frequency analysis; therefore, the peak heights and their statistical significance were calculated using the \textit{false alarm probability} and \textit{false alarm level} functions in \texttt{astropy}\footnote{\url{https://docs.astropy.org/en/stable/timeseries/lombscargle.html}}. The statistical framework of \citet{baluev08} was adopted for this purpose.

Frequency analysis was carried out separately for each survey and filter. Ultimately, we applied the same analysis to the entire data set, independent of filter and observatory. Figure~\ref{fig:freq_analysis} presents the resulting power spectra for all observatories, with horizontal lines indicating the 99\% confidence levels for each filter. The dominant frequencies were found at approximately 0.0003 d$^{-1}$ (3333~d) and its harmonic at 0.0006d$^{-1}$ (1667 d). Another prominent frequency was identified at $\sim$0.00229 d$^{-1}$ (437~days), denoted as $\omega$. Notably, we also detected its harmonic (218~d) in the T60 data. This signal was previously reported by \citet{kalomeni12} based on ROTSEIIId observations \citep{akerlof+03}.

\begin{figure}
\centering
\includegraphics[width=0.9\linewidth]{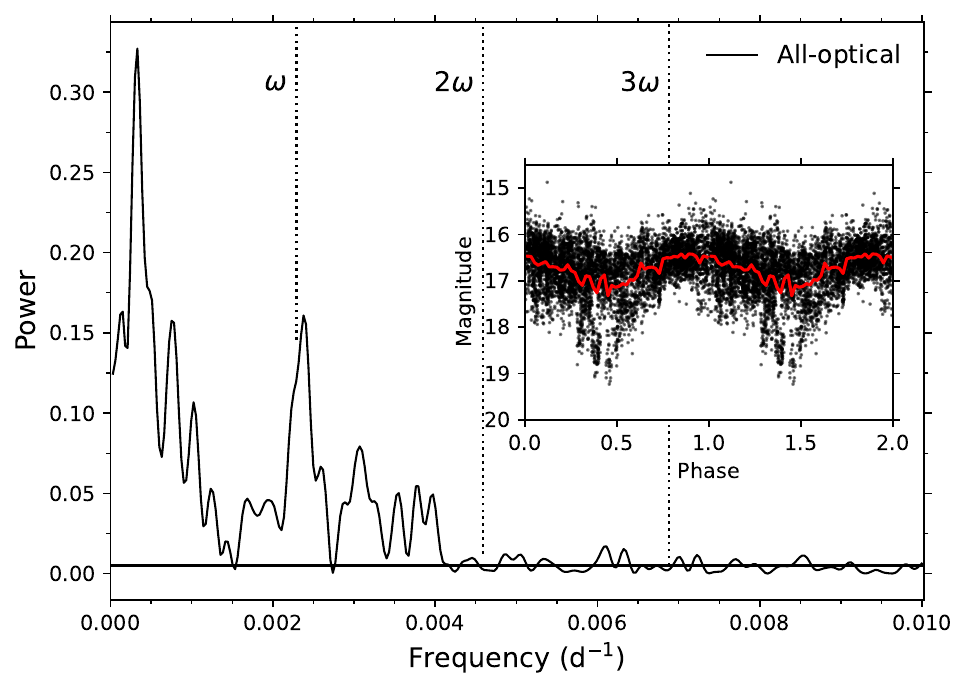}
\caption{The power spectrum of \textit{all-optical data} (1990 -- 2025) with an inset includes its folded light curve according to 0.00238(8)~d$^{-1}$. 
Dotted vertical lines show $\omega=0.00229$~d$^{-1}$ ($\approx$437~days) and locations of its harmonics. 
The horizontal line indicates 99\% confidence levels. The red line shows the grouped light curve into bins according to the 0.02 phase interval.}
    \label{fig:periodogram_218}
\end{figure}

To achieve a more comprehensive frequency analysis, we incorporated RoboScope data \citep[K\&H05~\textit{V};][]{kafka+05} with observations from other facilities, creating a unified data set representing 34 years of photometric monitoring of \anm. Frequency analysis was then performed on this combined data set (hereafter referred to as \textit{all-optical data}; see Fig.~\ref{fig:lc_ztf+atlas+crts}). The power spectrum of the \textit{all-optical data} is shown in Fig.~\ref{fig:periodogram_218}. The inset within the figure displays the phase-folded light curve corresponding to the detected frequency $f = 0.00238(8)$~d$^{-1}$.

Across most missions, a frequency close to $\omega = 0.00229$ d$^{-1}$ ($\approx$ 437 days) is evident. However, this signal is absent from the CRTS data set, which instead shows a variation at 280.75 d ($f = 0.00356(14)$ d$^{-1}$). The 437 days signal is more prominent in ASAS-SN, ATLAS, ZTF, K\&H05, and the \textit{all-optical data}. A common feature of these data sets is the absence of data brighter than 16 mag during their observational windows. By contrast, CRTS and AAVSO observations include periods of higher brightness and complex variability with amplitudes exceeding 2 mag, particularly between 1999 and 2010. It is therefore likely that this elevated and irregular brightness activity masked the 437~days signal in those data sets.

In addition, the period analysis was applied to the low accretion states where the deepest points are seen in Fig.~\ref{fig:lc_ztf+atlas+crts}, except RoboScope.
The intervals were selected as 2454935~<~BJD~<~2455155 and 2458900~<~BJD~<~2458960 for the first and second deeps, respectively. Unfortunately, AAVSO lacks sufficient data to determine the phase diagram based on the duration. The results for both deep structures are consistent with the orbital period of \anm.

\subsection{X-ray Photometry}

\begin{figure}
\centering
\includegraphics[width=0.98\hsize]{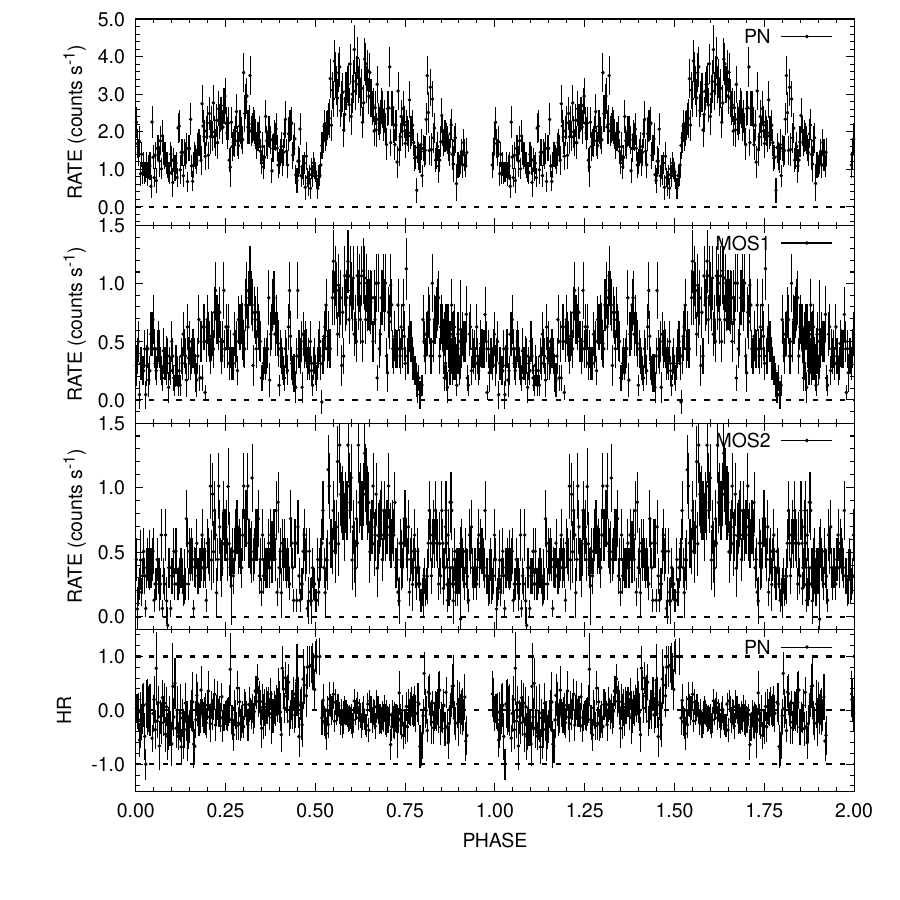}
\caption{\xmmn EPIC-pn, EPIC-MOS1 and EPIC-MOS2 light curves and Hardness Ratio of \anm.} 
\label{f:anuma_xray_lc}
\end{figure}

X-ray light curves and hardness ratio curves were constructed for the three cameras onboard \xmmn with the main purpose of understanding the accretion geometry and seeing the thermal behaviour of the main accretion region. The light curves were constructed in different energy bands as $0.2 - 10$~keV, $0.2 - 1$~keV, and $1 - 10$~keV with 20-second time bins to construct Hardness Ratio curves. Observations with EPIC-MOS cover a bit more than a full orbital cycle (1.04 cycles), while observations with EPIC-pn lasted just 0.95~orbital cycles. The folded light curves are shown in Fig.~\ref{f:anuma_xray_lc}.

All light curves exhibit a prominent X-ray hump at phase 0.63 and narrow dips around phases 0.48 and 0.80. The folded X-ray light curves provide the following information: the primary dip or eclipse-like feature occurs between phases 0.46 and 0.55 (centred at 0.48) with a duration of approximately 10 minutes. The X-ray flux begins to decline at phase 0.25 and continues until phase 0.5. A second narrow dip, more prominent in the EPIC-MOS data, begins at phase 0.77 and extends to 0.80 (centred at 0.785), lasting for around 3 minutes. X-ray counts remain positive across all phases, except during the eclipse-like dip, which is most apparent in soft X-rays.

To better understand these dip features and their spectral properties, we calculated the hardness ratio (HR) throughout the time series in specific energy bands. The hardness ratio, defined as \text{[(H-S)/(H+S)]}, was computed for the MOS and PN detectors using energy intervals of $0.2$–$1$~keV (soft) and $1$-$10$~keV (hard), with 20-second time bins. Variability in the HR is expected when X-ray emission is influenced by distinct absorption structures, such as accretion streams or curtains. The resulting variations in HR and energy-dependent light curves are shown in Fig.~\ref{f:anuma_xray_lc}. Apart from the primary and secondary dips, the HR generally fluctuates around $-0.1$. However, in the case of the main dip, EPIC-pn data clearly show a gradual increase in hardness before and during the dip, particularly in soft X-rays. This hardening effect is less evident in the MOS data due to lower signal-to-noise ratios. The behaviour of the secondary dip is especially intriguing: it is distinctly visible in the hard X-ray band, whereas the soft band shows a less pronounced variation.

Soft X-ray photons in polars are typically emitted from the magnetic poles of the white dwarf. Plasma channelled along magnetic field lines impacts these poles in free fall, generating hard X-rays in the post-shock region primarily through bremsstrahlung cooling. These photons are then reprocessed in the thermalised region, contributing to the observed soft X-ray emission \citep{lamb+79}. Accordingly, a correlation is expected between the accretion rate and the resultant X-ray photon flux \citep{hoshi73}.

Soft X-ray photons are highly susceptible to scattering when the emission region is obscured by accreting plasma. The accretion curtain or stream is typically responsible for intercepting soft X-ray photons, which leads to an increase in hardness and is generally accepted as the cause of the HR variations \citep{mukai17}. The secondary dip likely results from the scattering and absorption of hard X-rays, with two potential interpretations. First, the dip could arise from self-eclipse of the white dwarf during orbital motion. In this scenario, the soft X-ray region—covering a broader area of the accretion column—remains visible, while the more localised hard X-ray region becomes obscured. Second, depending on the intensity and geometry of accretion, the region may be temporarily concealed by an optically thick accretion structure, such as a stream or curtain.

When these eclipse-like features are considered alongside TESS observations, the accretion geometry becomes more apparent. The pronounced hump in the \xmmn light curve and the narrow absorption dip at phase $\sim$0.48, likely associated with the accretion stream or curtain, correspond closely to features observed in the optical light curves from TESS Sectors 21 and 75. The long-term light curve indicates that both sectors were observed during a low accretion state. Notably, in Sector 75, an eclipse-like feature linked to the accretion stream is present in isolation. As the mass accretion rate increases, the accretion stream is expected to become denser, eventually obscuring optical photons. A similar behaviour is observed in the well-studied polar V808~Aur, where the accretion stream becomes significantly more pronounced at higher accretion rates, just before the eclipse \citep{schwope+15, worpel+15}.

It is essential to consider the significance of hump-like structures, which are prominently observed in both optical and X-ray bands, when interpreting the system's geometry. Cyclotron emission is inherently beamed, producing two peaks or humps per orbital cycle due to the varying orientation of the emission region relative to the observer \citep{wickramasinghe+00}. It is, therefore, inappropriate to base conclusions about single- or double-pole accretion solely on optical photometry. Supporting data from spectroscopy and polarimetry are required. However, X-ray observations from \textit{EXOSAT}, \textit{ROSAT}, and more recently from \xmmn provide strong evidence for two-pole accretion. In particular, \textit{ROSAT} and \textit{EXOSAT} revealed a secondary hump between phases 0.85 and 1.25, which was not present in the \xmmn data \citep[see Fig.1 in][]{ramsay+94}. The two humps are separated by approximately 0.5 phase units—strongly supporting double-pole accretion. The isotropic nature of X-ray emission allows maximum flux when the emission region faces the observer. Furthermore, the phases of bright X-ray emission correspond to similar features in the optical data, particularly in TESS sector 48. This alignment suggests that each magnetic pole emits its own cyclotron radiation, with the secondary hump associated with the opposing pole. Taken together, these findings support the conclusion that the system time to time undergoes two-pole accretion.

\subsection{X-ray Spectroscopy}

\subsubsection{XMM-Newton Data}

The \xmmn spectrum of \anm was analysed using version 12.8.2 of the XSPEC package \citep{arnaud96,dorman+01}. Spectral fitting was performed for the EPIC-pn and EPIC-MOS instruments over the energy range of 0.2–10~keV. We initially fitted the data from both instruments using a single {\tt APEC} plasma model, assuming Solar abundances \citep{wilms+00} and a redshift fixed at zero \citep{mewe+85, liedahl+95}.

\begin{figure}
   \centering
   \includegraphics[width=1.04\linewidth]{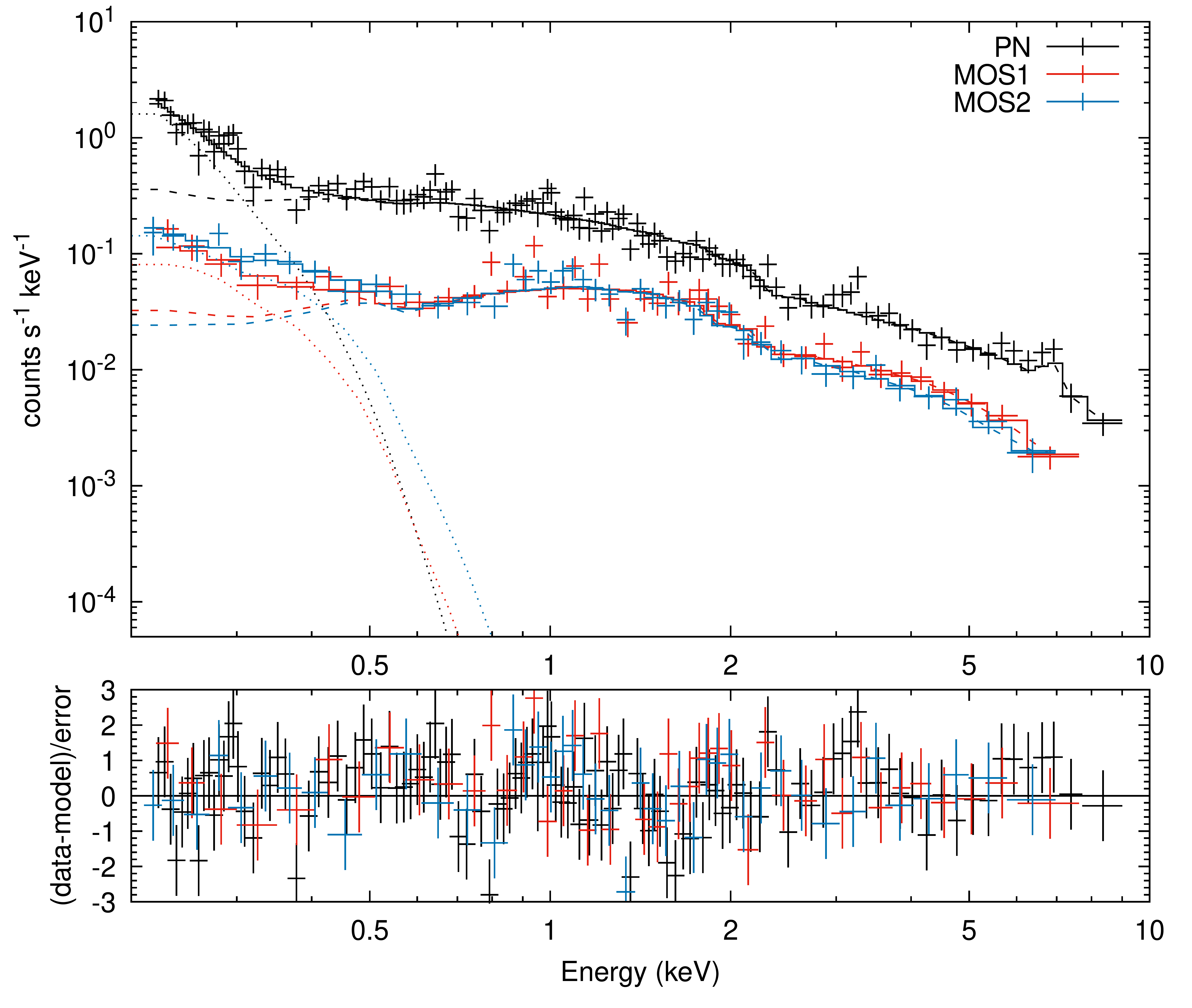}
      \caption{\xmmn spectra of \anm obtained from bright phases. 
 EPIC-pn spectra are grouped with 20 counts per bin. The lower panel shows residuals. For each camera, the dotted lines indicate the cold ({\tt BB}) component, and the dashed lines indicate the hot ({\tt APEC}) component.}
    \label{f:xmmspec}
\end{figure}

We defined the 0.60–0.75 phase interval as the bright phase based on the photometric light curves. This phase is assumed to represent the thermal state of the primary accretion region. A blackbody-like excess is evident in the softer part of the spectrum (see Fig.~\ref{f:xmmspec}). To model this feature, we constructed a baseline model comprising a blackbody ({\tt BB}) component to represent the soft thermal emission, and a hot thermal plasma {\tt APEC} model to characterise the emission from the hotter region (> 0.5 keV). This model configuration is commonly adopted for polars \citep{kuulkers+10}, and a similar setup was previously employed for the \textit{ROSAT} observations of \anm \citep{ramsay+94}. We subsequently included an absorption component to account for interstellar extinction, resulting in the model expression {\tt TBABS*(BBODY+APEC)} within {\tt XSPEC}. The neutral hydrogen column density in this direction is $N_{\rm H,gal} = 7.75\times10^{19}$,cm$^{-2}$ \citep{bekhti+16}. This model was applied to all spectra obtained with the EPIC-pn and EPIC-MOS instruments. The base model provided a good fit to the spectra and adequately reproduced the thermal emission within the effective energy range of \xmmn. The derived parameters, including temperatures, X-ray fluxes, and fit statistics, are summarised in Tab.~\ref{t:xray_spec_param}. In this table, we also report the joint fitting results for all three detectors. However, due to significant scatter in the MOS data—particularly within the 0.8–3 keV range—and a low signal-to-noise ratio, we adopted the EPIC-pn data as the reference for the subsequent analysis.

\begin{table*}
    \centering
    \caption{Fit parameters with their uncertainties, model bolometric fluxes, and luminosities of \anm. Spectra obtained during the bright phase via \xmmn.}
\begin{tabular}{lcccc}
     \hline
     \hline
    Model: TBABS*(BB+APEC) & PN &  MOS1 &  MOS2 & PN+MOS1+MOS2\\
     \hline
     \hline
     Parameters  &   &\\
     \hline
     N$_{\rm H}$ $(10^{22}$ cm$^{-2})$ & $<$ 0.008 
     & $<$ 0.008 & $<$ 0.008 & $<$ 0.008 \\
     kT$_{\rm bbody}$ (eV) & $30^{+4}_{-4}$
     & $29^{+10}_{-11}$ & $40^{+15}_{-10}$ & $34^{+4}_{-3}$ \\
     kT$_{\rm apec}$ (eV)  & $15^{+5}_{-4}$ 
     & $24^{+13}_{-13}$ & $10^{+11}_{-6}$ & $16^{+4}_{-5}$\\
     $\chi^2$ (d.o.f) & 1.04(119/114)
     & 1.13(45/40) & 1.13(44/39) & 1.03(202/197)\\
     \hline
     Unabsorbed Fluxes ($\times$10$^{-13}$ erg cm$^{-2}$ s$^{-1}$) &  \\
     \hline
     $F_{0.5-2.5}$ & $5.55^{+0.28}_{-0.30}$ 
     & $5.1^{+0.25}_{-0.28}$ & $4.9^{+0.2}_{-0.3}$ & $5.2^{+0.23}_{-0.12}$\\
     $F_{2.5-10}$ & $10.69^{+0.13}_{-0.13}$
     & $10.9^{+1.18}_{-1.45}$ & $8.6^{+0.6}_{-0.8}$ & $10.3^{+0.69}_{-0.8}$\\
     $F_{\rm bbody}$   & $58^{+12}_{-26}$
     & $75^{+10}_{-11}$ & $44^{+21}_{-13}$ & $29^{+9}_{-11}$ \\  
     $F_{\rm apec}$  & $28.53^{+2.4}_{-2.2}$ 
     & $35^{+10}_{-15}$ & 21 & $28^{+4}_{-5}$ \\  
     \hline
     $L_x$ ($10^{31}$ erg s$^{-1}$) & 5.3$\pm$0.5
      & 6.8$\pm$0.7 & 4.0$\pm$0.4 & 3.8$\pm$0.8\\
      \hline
      \hline
    \end{tabular}
    \label{t:xray_spec_param}
\end{table*}

The X-ray spectra of \anm reveal the presence of at least two emission components. Spectral fitting yielded a relatively low column density along the line of sight to the source, with values in the range of $(10^{-9} - 10^{-8}) \times 10^{22}$ cm$^{-2}$, significantly lower than the Galactic column density in this direction. Although the fitted $N_{\rm H}$ value is comparatively lower than the Galactic value, its inclusion notably influences the {\tt APEC} temperature: excluding it results in temperature variations of approximately $\pm$1–2 keV. Nevertheless, we argue that retaining $N_{\rm H}$ provides a more physically accurate representation of the absorbing environment. Therefore, we maintain this parameter in our model.

The X-ray fluxes for the {\tt BB} and {\tt APEC} components were computed separately across the $10^{-6}$–1000 keV energy range using the {\tt dummyrsp} command in {\tt XSPEC}. These fluxes were derived from the best-fitting parameters and applied to each model using the appropriate response matrix files (RMFs) defined for the \xmmn instruments via the {\tt flux} command.

To investigate the spectral characteristics during the primary dip, a spectrum was extracted for the 0.46–0.52 phase interval. In modelling this dip spectrum, the temperature parameters of the baseline model were held fixed, while the normalisation and absorption parameters were allowed to vary. The best-fitting model yielded a column density of approximately $N_{\rm H} \approx 2.4 \times 10^{21}$ cm$^{-2}$—a value approximately 31 times higher than the Galactic column density in this direction.

In polars, X-ray emission is generally associated with the instantaneous mass accretion rate ($F_{\rm acc} \sim F_{\rm X}$), a well-established relation in X-ray emitting systems. The total X-ray luminosity derived from the model flux offers a direct estimate of the mass accretion rate at the time of observation, based on the relation $\dot{M} \approx L_{\rm X} R_{\rm WD} / (G M_{\rm WD})$. Using this expression, and adopting a white dwarf mass of 0.8 \msun and a radius computed from the mass–radius relation of \citet{nauenberg72}, we estimated the X-ray luminosity of \anm’s primary accretion column as $L_{\rm X} = (5.3 \pm 0.5) \times 10^{31}$ erg s$^{-1}$. This corresponds to a mass accretion rate of $\dot{M} = (5.6 \pm 0.7) \times 10^{-12}$ \msunyr during the bright phase. Taking into account the fluxes from the EPIC-MOS1, EPIC-MOS2, and EPIC-pn instruments, we estimate the overall accretion rate to lie within the range $\dot{M} \sim (4$-$7) \times 10^{-12}$ \msunyr, assuming a distance of 334 pc and a white dwarf mass of 0.8 \msun.

\begin{figure}
\centering
\includegraphics[width=1.02\linewidth]{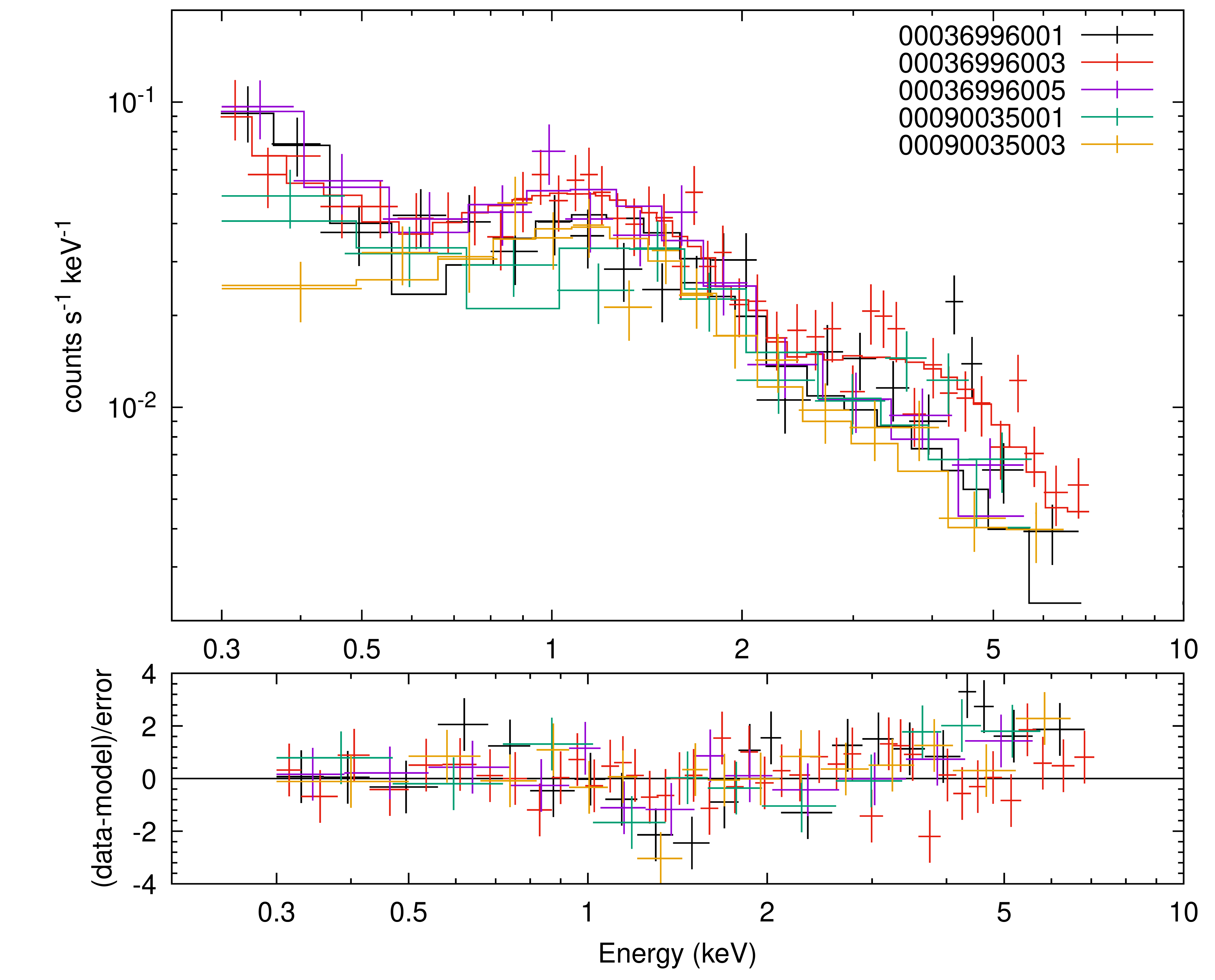}
    \caption{ Mean Swift/XRT spectra of AN UMa. The energy range covers 0.3 - 10 keV and each spectrum was created with 20 counts per bin. The lower panel shows residuals.} 
\label{fig:swiftspec}
\end{figure}

The blackbody temperature was determined to be approximately 30~eV across all instruments. Given that the effective lower energy limit of \xmmn is around 0.2 keV, the associated uncertainties at such low (soft X-ray) energies are well recognised in the study of polars \citep{mukai17}. The flux from this component can also be employed to estimate the current size of the accretion region on the white dwarf’s surface. Using the Gaia DR3 distance of 334~pc, we computed the blackbody-emitting area on the white dwarf to be $174 \pm 55$ km— a value consistent with those typically observed in polars. For comparison, \citet{schwope+20a} reported a similar accretion region radius of approximately 100km for the prototype polar AM~Herculis. Our estimate corresponds to approximately 0.15\% of the total surface area of a white dwarf with a mass of 0.8 \msun and a radius of $R_{\rm WD} \approx 6300$~km.

\subsubsection{Swift/XRT Data}

In this section, we present the X-ray observations of \anm obtained by \textit{Swift} between 2007 and 2009. Although 21 observations were conducted, \anm was detected in only five, and these were not suitable for phase-resolved spectral analysis. Except for one instance, all exposures were shorter than the orbital period, and the good time intervals were notably short. Nevertheless, the cumulative exposure time was sufficient to characterise the mean X-ray state of \anm and to trace its long-term X-ray behaviour in relation to accretion rate variability. The \textit{Swift} spectra, acquired in photon-counting mode, were modelled using {\tt XSPEC} within the $0.3$--$10$~keV energy range. Spectral analysis was conducted on spectra grouped to a minimum of 20 counts per bin, employing $\chi^2$ statistics.

The spectra from the five detections are shown in Fig.~\ref{fig:swiftspec}. All exhibit morphological similarities to the \xmmn spectrum. Consequently, we retained our primary spectral model, {\tt TBABS*(BB+APEC)}, under the assumption that the accretion column maintains similar physical properties, and modelled the data accordingly. For the observation with ID 90035003, no hump or soft excess consistent with a {\tt BBODY} component was evident below 2 keV. Therefore, in this case, we applied only a single {\tt APEC} model with an absorption component.

One of the principal challenges in fitting these spectra is the determination of the temperature of the hot plasma component ({\tt APEC}). The limited photon statistics and restricted energy coverage, common to all \textit{Swift} spectra, hindered reliable temperature determination. As such, we fixed the {\tt APEC} temperature at 15 keV, consistent with the value derived from \xmmn, assuming it to be representative of the \textit{Swift} observations. With this constraint, we allowed the remaining model parameters to vary in order to investigate the thermal evolution of the {\tt BBODY} component and the trend in absorption. The resulting best-fit parameters and temperatures are summarised in Tab.~\ref{t:swiftpar}.

\begin{table*}
    \centering
    \caption{Mean X-ray spectrum model parameters for Swift satellite. $F_{\rm apec}$ and $F_{\rm bbody}$ unabsorbed fluxes cover total energy interval of $10^{-6} - 10^{3}$~keV.}
     \begin{adjustbox}{width=\textwidth,keepaspectratio}
     \begin{tabular}{lcccccccccc}
     \hline
     \hline
     \multirow{2}{*}{Obs. ID} & BBODY & APEC & TBABS & PCFABS & PCFABS  & $F_{\rm bbody}\times10^{-11}$ & $F_{\rm apec}\times10^{-11}$ & $L_{\rm x}\times10^{32}$ & RATE & $\chi^{2}$ (dof)\\
     &  kT(eV)  & kT(keV) & ($10^{22}$cm$^{-2}$) & ($10^{22}$cm$^{-2}$) &  (CvrFract) &  (erg cm$^{-2}$ s$^{-1}$) & (erg cm$^{-2}$ s$^{-1}$) & (erg s$^{-1}$) &  ($0.3 - 10$ keV) & 
     \\
     \hline
     \hline
     36996001 & 34$\pm$6 & 15 & 0.11$\pm$0.06 & - & - & 20$\pm$1 & 1.0$\pm$ 0.3 & 13$\pm$2 & 0.11$\pm$ 0.005 & 1.22(22/18) \\
     36996003 & 34$\pm$12 & 15 & - & 12.7$\pm$4.5 & 0.59$\pm$0.05 & 2.9$\pm$0.3 & 2.7$\pm$0.6 & 3.6$\pm$0.3 & 0.131$\pm$ 0.004 & 1.13(43/38) \\
     36996005 & 36$\pm$15 & 15 & 0.02$\pm$0.01 & - & - & 2.3$\pm$0.6 & 1.0$\pm$0.4 & 1.64$\pm$0.16 & 0.11$\pm$ 0.006 & 1.44(13/9) \\
     90035001 & 46$\pm$9 & 15 & 0.43$\pm$0.13 & - & - & 38$\pm$90 & 0.96$\pm$0.27 & 10.2$\pm$0.8 & 0.09$\pm$ 0.006 & 1.57(11/7)\\
     90035003 & -  & 15 & 0.02$\pm$0.02 & - & - & - & 0.69$\pm$0.25 & 0.4$\pm$0.04 & 0.08$\pm$ 0.004 & 1.14(16/14) \\
     \hline
    \end{tabular}
    \end{adjustbox}
    \label{t:swiftpar}
\end{table*}

Following spectral fitting, we obtained consistent {\tt BBODY} temperatures across all observations. Despite differences in mass accretion rates, the derived temperatures agree within their respective uncertainties. In contrast, the column densities show significant variation. This dispersion may be attributed to the differing orbital phases and mass accretion levels at the time each observation was conducted. Using the same methodology applied to the \xmmn data, we calculated the total X-ray luminosity for each observation and determined the corresponding range in the mass accretion rate for \anm from the Swift data. Based on the lowest and highest X-ray flux levels, we find that the mass accretion rate ($\dot{M}$) in \anm varies between (1.35$\pm$0.15)$\times$10$^{-10}$\msunyr\ and (4.4$\pm$0.5)$\times$10$^{-12}$\msunyr.

\subsection{Spectral energy distribution}
\label{s:sed}

\begin{figure}
\centering
\includegraphics[width=8.9cm]{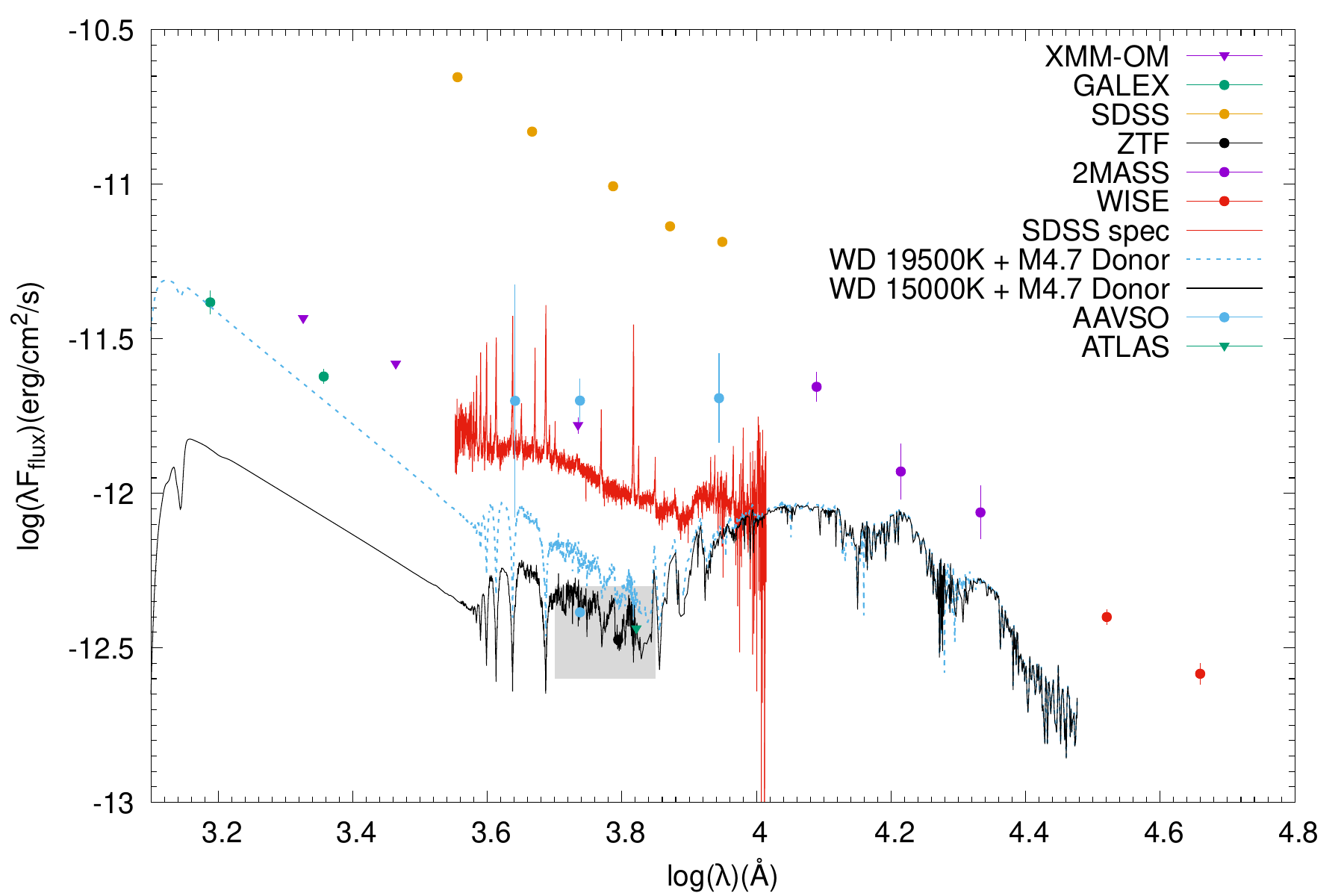}
\caption{Spectral energy distribution of \anm. Atmosphere models fit the observational points of GALEX (dashed blue line), ZTF, AAVSO, and ATLAS data points (black line) in the SED. Distance error of \anm is $\pm15$ pc.}
\label{f:sed}
\end{figure}

In Fig.~\ref{f:sed}, the spectral energy distribution of \anm is presented. The distribution has been constructed with archival data from sky survey programs;  Wide-field Infrared Survey Explorer (WISE), the 2-Micron All-Sky Survey (2MASS), the Sloan Digital Sky Survey (SDSS), the Pan-STARRS survey, GALEX, \xmmn optical monitor, AAVSO, and TUG T60, using the effective wavelengths and zero points for each filter given in Spanish Virtual Observatory\footnote{\url{http://svo2.cab.inta-csic.es/theory/newov2/}} (SVO) \citep{bayo+08}. The effective wavelengths and zero points for the \xmmn OM filters are given in \cite{kirch+04}. The SDSS spectrum of \anm obtained on the way of low state in March 2013 was also added to the distribution. In SDSS spectrum, the mean spectral flux density is about 1.9$\times$10$^{-16}$ erg cm$^{-2}$ s$^{-1}$ \mbox{\,\AA}$^{-1}$ at 5500\,\AA\, which corresponds to Johnson V-band magnitude of 18.2 (Vega). This shows that the observation was made in an increased low-state during the observation. However, SDSS brightnesses were obtained at a high state. Synthetic atmosphere models on the SVO database were used to describe the best fit to observational data points. \cite{koester+10} models were used to determine white dwarf atmosphere, while NextGen's synthetic models were utilised for late-type donor star \citep{allard+97,baraffe+97a,baraffe+98,hauschildt+99}. We estimated the upper limit for the physical parameters of binary components of the system using Gaia distance and the semi-empirical sequence derived by \cite{knigge+11}. We chose a white dwarf mass of 0.8~\msun, considered the average mass for cataclysmic binaries \citep{pala+21}. The mass-radius relation of the WD was calculated from \cite{nauenberg72}. 

In fitting the synthetic models, we utilised the ultraviolet magnitudes obtained from \xmmn/OM and the faintest optical brightness levels recorded in ATLAS, ZTF, and AAVSO observations. As shown in Fig.~\ref{fig:lc_ztf+atlas+crts}, two deep minima are evident, during which \anm exhibits a pronounced decline in brightness. These low-state parts are characterised by a near-complete cessation, or substantial reduction, in mass transfer, whereby the radiative output from the white dwarf and the donor star becomes dominant.

The magnitudes at minimum brightness (BJD$_{\rm min}$, mag) observed during these low mass accretion states are approximately as follows: ($\sim$2449408, 18\fm22; $\sim$2452263, 18\fm26) from RoboScope observations \citep{kafka+05}; (2454960.83, 19\fm24) from AAVSO-\textit{V}; (2455005.66, 17\fm93) from CRTS-\textit{V}; (2458942.94, 19\fm10 and 2460451.33, 18\fm93) from ATLAS-\textit{o}; and (2458939.87, 19\fm23) from ZTF-\textit{r}.

The ultraviolet brightness is particularly informative, as emission in this regime primarily originates from the atmosphere of the white dwarf. For a white dwarf mass of 0.8~\msun, we find that a 19500 K atmosphere model provides a good match to the observed ultraviolet flux. However, when this model is extended into the optical range, it overestimates the flux at the low-state brightness levels recorded in ATLAS, ZTF, and AAVSO, even when combined with an M4.7-type donor star model (see dashed blue line in Fig.~\ref{f:sed}). Regardless of the assumed temperature of the secondary star, the combined synthetic model consistently exceeds the observed optical fluxes, primarily due to the white dwarf's optical contribution.

This discrepancy implies that the ultraviolet measurements may include additional emission components associated with residual accretion processes. Therefore, we suggest that the actual temperature of the white dwarf must be lower than 19500~K to match the observed optical data. Taking into account the upper uncertainties in the ATLAS and ZTF measurements, we find that a 15000~K white dwarf model provides a good fit to the optical data. The uncertainty in distance contributes a temperature error of $\approx$ 250~K.

Unfortunately, in the red part of the SED, no data corresponding to the low-accretion state are available that can be reliably attributed to the donor star’s atmosphere. Data in this region are contaminated by cyclotron emission linked to ongoing accretion onto the white dwarf. Nevertheless, based on the system's orbital period of 0.07975 days, we estimate that the companion is likely to be of spectral type M4.7 (corresponding to a mass of $\approx$0.15\msun), consistent with the semi-empirical donor sequence proposed by \cite{knigge+11}. This value also lies below the upper mass limit ($\approx0.2$~\msun) for this period in the donor star evolution models of \cite{kalomeni+16}.

In conclusion, based on the available data, we propose that \anm likely contains a white dwarf with a temperature of approximately 15000 K and a mass of 0.8\msun, accompanied by a donor star of spectral type M4.7. These values should be regarded as upper limits, constrained by the system's low-state spectral energy distribution.

\section{CONCLUSIONS}
\label{sec:conclusions}
We have conducted a comprehensive investigation of the prototype polar AN~Ursae~Majoris (\anm), utilising all available data from \xmmn and \textit{Swift}, complemented by additional photometric observations from both ground- and space-based facilities. The data from the ATLAS survey revealed previously unreported low accretion states. Optical photometry from TESS shows that, during these low states, the accretion curtain occasionally becomes visible. Notably, ATLAS data also indicate that the accretion column undergoes a change during transitions between accretion states. Observations with the TUG T100 telescope demonstrate that, over certain intervals, accretion onto the secondary magnetic pole can become sufficiently strong for its brightness and visibility to rival that of the primary accretion column. Of particular significance are the optical light curves presented in \citet{bonnet-bidaud+96} (see their Fig.~2), which, when folded using a consistent ephemeris, exhibit a single-hump structure. This hump is centred around phase 0.0 and coincides with the location of the secondary magnetic pole. This provides compelling evidence that, at times, accretion in \anm occurs exclusively onto the secondary pole.

TESS photometry reveals three distinct accretion states in \anm, indicating changes in accretion geometry. Sector~48, obtained during a high state, exhibits a markedly different morphology, suggesting episodic accretion onto the secondary pole. X-ray counts (\xmmn) are minimal between phases 0.9 and 1.1, suggesting a self-eclipse of the primary accretion column by the white dwarf. Within this phase range, a small hump-like feature is detected in sector 48.  The centres of the primary and secondary humps are separated by 0.5 phase units. The visibility of the second hump covers 0.23 phase units, while the primary hump is visible over 0.77 phase units. A similar two-peaked profile in \textit{EXOSAT} and \textit{ROSAT} light curves supports transient two-pole accretion. Due to its isotropic nature, these two peaks are expected to originate from two different emission regions. The observation of optical cyclotron humps in the same phases supports two-pole accretion. Evidence of this behaviour also appears sporadically in long-term optical light curves and nightly TUG T100 observations (see Fig.~\ref{fig:atlas_8part}). The folded ATLAS data show that the amplitude of the mini-hump between phases 0.9 and 1.15 sometimes increases, and its visibility can extend up to 0.5 phase units. In contrast, Sectors 21 and 75, observed during low states, display reflective symmetry consistent with stable, single-pole accretion, corroborated by contemporaneous \xmmn data. These states coincide with faint low states and their transitions in ATLAS photometry. Sector~75 also shows an eclipse-like dip near phase 0.43, attributed to obscuration of the primary pole by an accretion curtain or stream, visible in both X-ray and optical bands. A double hump feature is seen in the majority of folded ATLAS and TUG observations concerning orbital period. These humps are separated by 0.5~phase units. In this context, it may be asserted that double hump or double pole mass accretion is typically observed in the high accretion state. At the poles, not only periodic but also cycle-to-cycle mass accretion flow behaviour to the other pole can be observed. An example is V496~UMa system, which is a polar that exhibits this behaviour due to its interesting pole locations and its variable nature \citep{ok+22,kennedy+22}. 

One of the most noteworthy findings from the light curves is that mass accretion onto the secondary magnetic pole can occasionally become dominant, with corresponding cyclotron-related optical humps clearly visible in the TUG T100, T60, and ATLAS observations. The strength of this accretion flow suggests that the magnetic poles are situated close to the white dwarf's rotational axis. In this context, the result of $\beta = 20^\circ$ for the colatitude of the accretion column, as measured by \cite{cropper+89}, appears to be consistent. In cataclysmic variables, accretion is generally expected to occur onto the more weakly magnetised pole. A comparison with other well-studied polars, such as VV~Pup \citep{wickramsinghe+89} and UZ~For \citep{schwope+90}, shows that accretion typically favours the less strongly magnetised pole. In the case of \anm, the cyclotron emission from the secondary pole reaches a brightness comparable to that of the primary in optical photometry, which is particularly intriguing. The magnetic field strength at the secondary pole could, in principle, be determined from cyclotron harmonic features observed in the spectrum, provided that accretion is occurring simultaneously onto both poles. This emission is highly sensitive to both magnetic field strength and the system's geometry \citep{schwope+90,kalomeni+05}. Nonetheless, based solely on photometric data, it may be inferred that the magnetic field strength at the secondary pole is likely comparable to that of the primary pole.

Our period analysis, based on TESS photometry, yielded a refined orbital period of 0.079752867(12) days, which is significantly more precise than the previous determination by \cite{bonnet-bidaud+96}. The difference between the two periods, $\Delta P \approx 4.2 \times 10^{-8}$ days, is within the uncertainty range, confirming that \anm remains synchronous within observational limits. The improved period accuracy strengthens the ephemeris for future observational campaigns and provides a reliable reference for detecting potential long-term period variations in the polars. ATLAS observations indicate that the locations of the dip structures, arising from a particular accretion geometry, are displaced when the geometry changes. In cases of asynchronism, it is anticipated that the dip positions will likewise vary with time, resulting in the observation of humps at distinct phases. Nevertheless, in the returns to the accretion geometry where the dip structures are observed, they reappear in the identical phase intervals in the light curves.

We confirm that the X-ray spectrum of \anm exhibits strong soft X-ray emission in both \xmmn and \textit{Swift} observations. The \textit{Swift} spectra yield comparable blackbody temperatures across all observations ($\approx$ 30 – 35 keV), with the exception of the final low-state X-ray observation. The absence of a detectable {\tt BBODY} component in this last \textit{Swift}/XRT (0.3 – 10 keV) spectrum may be attributed to the blackbody temperature lying outside the effective energy range of the \textit{Swift} observatory. The significant variability observed in soft X-ray photon flux is likely due to changes in the accretion flow, combined with variable photoelectric absorption \citep{watson+89}. Accordingly, the lack of a {\tt BBODY} component in the most recent \textit{Swift} observation may reflect a period of minimal accretion, consistent with this interpretation.

The \xmmn observations clearly reveal that the initial narrow dip occurring at phase 0.48 is associated with the accretion stream or curtain. The \textit{hardness ratio} remains largely unchanged throughout the remainder of the light curve, indicating no significant variation in the spectral shape. It is worth highlighting here about \textit{EXOSAT} \citep{bonnet-bidaud+96} and \textit{ROSAT} \citep{ramsay+94} observations. Both exhibited a two-hump structure not present in the \xmmn data. In light curves constructed using comparable $T_{0}$ and orbital periods, both \textit{ROSAT} and \textit{EXOSAT} display an additional minor hump near phase 0.0, which is absent in the \xmmn observations. In those earlier data sets, dip features—attributed to the obscuration of the accretion region—are located within this small hump. In contrast, the \xmmn light curve places these dip structures within the second hump, centred around phase 0.5. This suggests that dip features may appear at different orbital phases, forming absorption-induced substructures that can also influence the optical light curves. While the self-eclipse and absorption-related dips around phase 0.48 in the \xmmn data exhibit similar behaviour to those seen in TESS Sectors 21 and 75, the more complex variability pattern observed in TESS Sector 48, as well as the multiple dip structures detected in the corresponding ATLAS data, likely originates from the secondary magnetic pole and its associated absorption components.

The first discovered prototypical polar AM~Her frequently exhibits abrupt brightness fluctuations alongside extended intervals of low or high mass accretion states \citep{kafka+05}. In contrast, \anm displays markedly different brightness behaviour than AM Her. Transitions from low to high mass accretion states occur gradually rather than suddenly. On occasion, the system enters a very faint low state, with the brightness dropping to $\approx$19 mag. As noted by \citet{kafka+05}, such faint states may be difficult to detect due to observational limitations. Over the past 34 years, \anm has experienced seven short-lived low states, each reaching a magnitude fainter than $\leq$18. Notably, the minima observed in 1994 and 2020 exhibit strikingly similar features in both their fading and recovery phases. \citet{garnavich+88} reported variability in the range of $16-19$ mag for \anm but recorded only a single instance of the system at around 14.5 mag. Our findings indicate that \anm typically remains within a stable brightness range of 16–17 mag during high accretion state, with excursions beyond 16 mag being relatively rare. However, during the period between 1999 and 2009, observations from RoboScope, CRTS, AAVSO-\textit{V}, and AAVSO-\textit{R} reveal that \anm entered an exceptionally high accretion state, with brightness surpassing 16 mag.

The mechanism responsible for initiating mass accretion in cataclysmic variables remains poorly understood. It is commonly hypothesised that the magnetic activity of the donor star plays a crucial role in modulating state transitions \citep{kafka+05}. \citet{hessman+00} proposed that star spots on the donor's surface, particularly near the L1 point, may disrupt or modulate the mass transfer rate if they migrate across the region. However, their analysis focused solely on AM Her, a system characterised by relatively regular transitions between accretion states. This scenario appears less applicable to \anm, which exhibits sporadic and abrupt transitions into low states of very short duration. Unlike AM Her, \anm predominantly resides in a sustained high accretion state. While large fluctuations in brightness due to changes in the accretion rate are observed, the system’s low-state magnitude does not show stable, long, low states typical of AM~Her.

From the long-term light curve of \anm, we identify two periodicities: 437 and approximately 1667 days (along with a harmonic at 3334 days). The longer period aligns with timescales associated with activity cycles of M-type stars \citep[$\sim$5-8 years;][]{mascareno+16, ramsay+24}. However, the origin of the 437-day period remains ambiguous. This period is too long to correspond to the rotational period of the donor star and too short to represent a magnetic activity cycle. While this periodicity appears to correlate with transitions into low states, further extended monitoring is required to confirm its persistence. It is also important to consider whether such periodicities may arise from binary-related effects, although distinguishing these signals is challenging due to the abrupt brightness changes, which diminish frequency amplitudes in periodograms.

Historically, the distance to \anm has been a subject of considerable debate. With the advent of high-precision parallax measurements from \textit{Gaia} DR3, we now possess the most reliable distance estimate to date. Coupled with multi-band photometry and synthetic atmosphere models, these data have enabled us to place upper limits on key system parameters. Based on the faintest states observed in ZTF and ATLAS surveys, we infer that the system likely hosts a white dwarf with a temperature of 15000K and a mass of 0.8\msun, accompanied by a secondary star of spectral type M4.7 or later, corresponding to $\approx$ 0.15~\msun.

\anm stands out as a particularly unique polar, being among the earliest discovered examples of its class and supported by a rich archive of observational data. Our findings, derived from time-resolved X-ray and optical observations, offer new insights into the accretion geometry and variability of this system. Nevertheless, continued long-term photometric monitoring and coordinated X-ray observations will be essential to determine whether the observed characteristics persist or evolve over time, especially in light of potentially significant changes in mass transfer behaviour across different accretion states.

\section*{Acknowledgments}
This work was supported by the Scientific and Technological Research Council of T\"urkiye (T\"UB\.ITAK) under grant 123F148. SO gratefully acknowledge support for this project by TUBITAK 2214-A International Doctoral Research Fellowship Programme and T\"UB\.ITAK 2211-C Ph.D. Scholarship Programme. This article includes a part of the PhD thesis of SO. The authors thank T\"UB\.ITAK for partial support in using the T100 telescope (16BT100-1027) and T60 telescope (16BT60-1008). We acknowledge with thanks the variable star observations from the AAVSO International Database contributed by observers worldwide and used in this research. This work has made use of data from the European Space Agency (ESA) mission Gaia (https://www.cosmos.esa.int/gaia), processed by the Gaia Data Processing and Analysis Consortium (DPAC, https://www.cosmos.esa.int/web/gaia/dpac/consortium). This research has made use of data, software and/or web tools obtained from the High Energy Astrophysics Science Archive Research Center (HEASARC), a service of the Astrophysics Science Division at NASA/GSFC and of the Smithsonian Astrophysical Observatory’s High Energy Astrophysics Division. Based on data from the Spectral Stellar Libraries services developed by the Spanish Virtual Observatory in the framework of the IAU Commission G5 Working Group: Spectral Stellar Libraries. We are grateful to the anonymous referee, whose comments led to improvements in the clarity of the paper.

\section*{Data Availability}

TUG T100 and TUG T60 data underlying this article will be shared on reasonable request to the corresponding author. The other data used in this article are publicly available from the sources in the text.

\bibliography{anuma_rev4}{}
\bibliographystyle{aasjournal}

\label{lastpage}
\end{document}